\documentclass[a4paper,10pt]{article}
\pdfoutput=1

\usepackage[T1]{fontenc}
\usepackage{adjustbox}
\usepackage{tcolorbox}

\usepackage{graphicx}
\usepackage{pdfpages}
\usepackage{enumitem}
\usepackage{amssymb,amsmath,amsfonts,amsthm,titling,url,array,mathrsfs}
\usepackage{mathtools}
\usepackage{tikz}
\usetikzlibrary{calc}

\usepackage{xcolor}
\usepackage{latexsym}

\usepackage{jheppub}

\newcommand{\dd}{\text{d}}
\newcommand{\cc}{\text{c}}
\newcommand{\geo}{\text{g}}

\title{The Black Hole Entropy Distance Conjecture and Black Hole Evaporation}

\author[a]{Marvin L\"uben,}
\author[a,b]{Dieter L\"ust,}
\author[b]{Ariadna Ribes Metidieri}

\affiliation[a]{Max-Planck-Institut f{\"u}r Physik (Werner-Heisenberg-Institut),\\
F{\"o}hringer Ring 6, 80805 M{\"u}nchen, Germany}
\affiliation[b]{Arnold-Sommerfeld-Center for Theoretical Physics, Ludwig-Maximilians-Universit{\"a}t,\\
Theresienstr. 37, 80333 M{\"u}nchen, Germany}

\emailAdd{mlueben@mpp.mpg.de}
\emailAdd{dieter.luest@lmu.de}
\emailAdd{a.ribes@physik.lmu.de}

\abstract{We extend the recently proposed Black Hole Entropy Distance Conjecture to the case of charged black holes in de Sitter space.
By systematically studying distances in the space of black hole geometries with multiple horizons,
we find that the distance is generically related to the logarithm of the entropy. From the infinite distance conjecture
this predicts the appearance of a massless tower of modes in the limit of infinite entropy.
Further, we study the evaporation of these black holes and relate it to the geometric distance.
We find that the corresponding distance to the final stage of evaporation is finite.
We conclude that evaporation does not lead to the appearance of a light tower of black hole microstates.
}

\begin{document} 

\begin{flushright} 
\texttt{LMU-ASC 46/20, MPP-2020-210}
\end{flushright}

\setcounter{tocdepth}{2}
\maketitle

\section{Introduction}
\label{sec:intro}

The swampland program aims to distinguish those low-energy effective field theories that can be completed into quantum gravity in the ultraviolet from those that cannot~\cite{Vafa:2005ui}.
In order to decide whether an effective theory is in the swampland or not, a series of criteria have been conjectured.
For reviews we refer to~\cite{Brennan:2017rbf,Palti:2019pca}.
The Distance Conjecture~\cite{Ooguri:2006in} is of particular interest for the present paper.
It states that at an infinite distance in the moduli space of theories an infinite tower of states becomes massless.
Therefore, the effective theory is rendered invalid in the infinite distance limit.

While the distance conjecture was postulated for moduli, it was generalized to arbitrary fields in~\cite{Lust:2019zwm}.
Therein, the generalized distance conjecture was applied to background metrics within the context of Anti-de Sitter (AdS) space-times.
It was found that the flat limit is at infinite distance in field space.
This implies that the near-flat limit of pure AdS is accompanied by an infinite tower of light states.
The distance between background metrics was further applied to space-times with a single horizon in~\cite{Bonnefoy:2019nzv}.
The analysis shows that the limit of infinite entropy $\mathcal S$ is at infinite distance in field space.
Combined with the distance conjecture~\cite{Ooguri:2006in} it follows that a tower of modes becomes massless as
\begin{equation}
    m\sim \mathcal S^{-c}
\end{equation}
with $c\sim\mathcal O(1)$.
The limit $\mathcal S\to \infty$ is identified with Minkowski space-time.
This observation prompted the postulation of the Black Hole Entropy Distance Conjecture (BHEDC)~\cite{Bonnefoy:2019nzv}.
While in these studies the variation of the background metric was related to the geodesic flow, also other types of flow have been studied in the context of the swampland~\cite{Kehagias:2019akr,DeBiasio:2020xkv}.
The conjectures on AdS were explicitly tested in concrete string theoretic setups, e.g. in~\cite{Blumenhagen:2019vgj}.

Here we generalize~\cite{Lust:2019zwm,Bonnefoy:2019nzv} to spherically symmetric and static space-times with multiple horizons.
In particular, we consider effective theories with positive cosmological constant.
Therefore, we study metrics that belong to the family of Reissner-Nordstr{\o}m-de Sitter (RNdS) solutions.
Working with multi-horizon space-times that are asymptotically de Sitter poses some challenges such as the definition of the entropy.
For a space-time with a single horizon, the entropy of the space-time can be identified with the entropy associated to the single horizon.
The entropy is given by the area law~\cite{Bekenstein:1972tm,Hawking:1974sw,Gibbons:1977mu}
\begin{equation}
    \mathcal S = A/4\,,
\end{equation}
where $A$ is the area of the horizon in Planck units.
We will briefly discuss how the entropy can be defined for multi-horizon space-times.
On the other hand, within string theory the microscopic origin of the entropy is best understood for supersymmetric black holes~\cite{Strominger:1996sh}, but recently progress was made for Schwarzschild geometries in the context of soft hairs~\cite{Hawking:2015qqa,Hawking:2016msc,Averin:2016ybl,Averin:2016hhm,Hawking:2016sgy}.

Our strategy for generalizing the BHEDC is as follows.
Using the definition of geodesic distance between space-times~\cite{DeWitt:1984cck,Gil:1991,Lust:2019zwm}, we compute the distance between points in the moduli space of metrics.
In particular, we identify the infinite distance points.
We then relate the geodesic distances between points in moduli space to the entropy of these points.
As we will see, the space-time configuration with infinite entropy is at infinite geodesic distance.
On the other hand, all space-time configurations with finite entropy are at finite distance from each other.
Moreover, our study shows that the distance $\Delta$ is generically proportional to the logarithm of the entropy of the space-time,
\begin{equation}
    \Delta \sim \log\mathcal S\,.
\end{equation}
In other words, the BHEDC can be generalized to multi-horizon space-times.

This generalization allows to apply the BHEDC to physical phenomena in our universe.
In this paper, we study the evaporation of charged black holes in asymptotically de Sitter space-time due to Hawking and Schwinger.
This is the most general spherically-symmetric and static configuration that respects the completeness conjecture~\cite{Polchinski:2003bq} and the no hair theorems~\cite{Israel:1967wq,Israel:1967za}.
Further, in~\cite{Montero:2019ekk,Antoniadis:2020xso} RNdS black holes were used to formulate de Sitter versions of the Weak Gravity Conjecture~\cite{ArkaniHamed:2006dz}.
We will take the conjecture of~\cite{Montero:2019ekk} as a basis such that the black hole evaporates quasi-statically.
In addition, we consider the cosmological constant to vary in time to circumvent the de Sitter conjectures~\cite{Dvali:2014gua,Dvali:2017eba,Obied:2018sgi,Dvali:2018fqu,Garg:2018reu,Ooguri:2018wrx,Dvali:2018jhn}.
We implement the variation of the cosmological constant via thermodynamic considerations~\cite{Dolan:2013ft}.
The Hawking and Schwinger effects accumulate and backreact on the geometry.
This can be interpreted as a variation of the background metric, which travels a distance in the moduli space.
Hence, we can use our generalized prescription to compute the geometric distance between the initial and final point of the evaporation process and relate the inferred distance to the Generalized Distance Conjecture.
As we will see later, evaporation leads to a finite distance in moduli space.
Hence, this toy model of self-similar and quasi-static evaporation satisfies all these swampland criteria.

The organization of this work is as follows:
in Sec.~\ref{sec:RNdS} we introduce the Reissner-Nordstr\o m-de Sitter solution of Einstein field equations.
The reader familiar with the RNdS solution can safely skip this section and only use it as a reference for the notation and unit system used. 
Sec.~\ref{sec:ModuliSpaceAndGeodesics} is devoted to the construction of the moduli space of metrics, the solution of the geodesic equation and the computation of moduli space distances between spherically symmetric and static manifolds.
Next, we discuss how to define entropy in the space-times with multiple horizons in Sec.~\ref{sec:BHEDC}.
Further, we establish a relation between moduli space distances and entropy.
In Sec.~\ref{sec:BHEvaporation}, we analyze the evaporation of Reissner-Nordstr{\o}m and Schwarzschild black holes in asymptotically de Sitter backgrounds and identify the final stage of evaporation.
We compute the distances between the initial and final space-times of the evaporation process.
Finally, we conclude in Sec.~\ref{sec:Conclusion}.

\section{Reissner-Nordstr\o m black holes in de Sitter}
\label{sec:RNdS}
In this section we review the black hole solution
of Einstein gravity with cosmological constant coupled to Maxwell's theory.
We work in (3+1)-dimensional space-time and use the
$(-,+,+,+)$-metric signature.
The action for the metric tensor $g_{\mu\nu}$ and
gauge field $A_\mu$ is given by~\cite{Reissner:1916,1918KNAB...20.1238N}
\begin{equation*}
S = \int_M \dd^4x\sqrt{-g} \left[\frac{1}{16 \pi G} (R-2\Lambda) - \frac{1}{4g_0^2}F_{\mu\nu} F^{\mu\nu}\right]
\end{equation*}
where $G$ is Newton's constant,
$R$ is the Ricci scalar, 
and $\Lambda$ the cosmological constant.
Further, $F_{\mu\nu}$ is the electromagnetic field strength tensor and $g_0$ the $U(1)$ gauge coupling.

In Schwarzschild coordinates $(t,r,\theta,\phi)$
the most general, spherically symmetric, and static solution
for the metric $g_{\mu\nu}$ and gauge field $A_\mu$
is given by
\begin{equation}\label{metricRNdS}
\dd s^2 = - V(r) \dd t^2 + V(r)^{-1} \dd r^2 +r^2 \dd S_2^2\,, \qquad A = -\frac{g_0^2}{4\pi}\frac{q}{r} \dd t
\end{equation}
with $\dd S_2^2 = \dd\theta^2+\sin^2\theta \dd\phi ^2$
the metric on the 2-sphere and
\begin{equation} \label{Vsubextremal}
V(r) = 1-\frac{2Gm}{r} +\frac{g_0^2 G}{4\pi}\frac{q^2}{r^2} -\frac{r^2}{\ell^2}\,.
\end{equation}
The parameters $m$ and $q$ are related to the mass and charge
of the space-time~\cite{Sekiwa:2006qj,Dolan:2013ft},
and we will refer to $\ell = \sqrt{3/\Lambda}$
as de Sitter radius with $\Lambda>0$.
This is the Reissner-Nordstr\o m-de Sitter (RNdS) solution
that describes an electrically charged black hole in an asymptotic de Sitter space-time.
For the sake of brevity, we define the mass parameters
\begin{equation}\label{eq:dimless_mass_par}
M = Gm\,, \quad
Q = q\sqrt{\frac{g_0^2G}{4\pi}}\,.
\end{equation}
The metric function then takes the compact form
\begin{equation}\label{eq:V_norm}
V(r) = 1-\frac{2M}{r} +\frac{Q^2}{r^2} -\frac{ r^2}{\ell^2}\,.
\end{equation}

The causal structure of RNdS space-time is characterized by three horizons.
A causal horizon can be defined as the null hyper-surface where the metric changes its signature.
Since the RNdS space-time is static, the Killing, apparent and event horizons coincide~\cite{poisson_2004}.
Further, for spherically symmetric space-times the aforementioned horizons are located at the radial position $r=r_h$ where the metric function \eqref{eq:V_norm} vanishes,
\begin{equation}\label{eq:poly_eq}
    V(r)\Big|_{r=r_h} = 0 \,,
\end{equation}
which represents a quartic polynomial.
The number of real roots is dictated by the sign of the discriminant locus $D$ of that quartic polynomial, explicitly given by 
\begin{equation} \label{delta}
\frac{D}{16} = \frac{M^2}{\ell^2}-\frac{Q^2}{\ell^2}-\frac{27M^4}{\ell^4}+\frac{36M^2Q^2}{\ell^4}-\frac{8Q^4}{\ell^4}-\frac{16Q^6}{\ell^6}\,.
\end{equation}
For $D\geq0$, Eq.~(\ref{eq:poly_eq}) has a maximum of four real-valued roots, but one of them is always negative and hence non-physical.
The space-time can have a maximum of three causal horizons.
These are the inner black hole horizon $r_-$, the outer black hole horizon $r_+$ and the cosmological horizon $r_\text{c}$.
Only the horizons $r_+$ and $r_\cc$ are accessible to an observer outside of the black hole.
The explicit expressions for the horizons $r_{\{\pm,c\}}$ as functions of the parameters (\ref{eq:dimless_mass_par}) can be found in Appendix \ref{sec:appA}.

Next let us define the phase space of RNdS solutions as the 3-dimensional parameter space spanned by the mass $M$, charge $Q$ and de Sitter radius $\ell$.
To respect the Cosmic Censorship Conjecture~\cite{Penrose:1969pc}, i.e. to ensure absence of naked singularities\footnote{See, e.g.~\cite{Virbhadra:1998dy,Virbhadra:2002ju,Virbhadra:2007kw,Xu:2019wak} for some related work on naked singularities.}, we impose $M\geq 0$ and $D\geq 0$.
The latter condition implies that all three horizons are real and satisfy $r_\text{c}\geq r_+ \geq r_-$.
Further, we impose $\ell\geq 0$ to exclude Anti-de Sitter.
We refer to the region that respects these conditions as \textit{physical phase space} $\mathcal D$.

The boundary of the physical phase space, that we denote by $\partial \mathcal{D}$, is characterized by $D=0$.
This implies that two or three horizons are degenerate.
We distinguish three such cases.
The case where the black hole and cosmological horizons coincide, $r_+ = r_\cc$, is referred to as Nariai solution.
If instead the inner and outer black hole horizons coincide, $r_-=r_+$, the space-time is called extremal.
Finally, the space-time with all three causal horizons being degenerate, $r_\cc=r_+=r_-$, is called ultra-cold.

\begin{figure}
\centering
\includegraphics[width=0.8\textwidth]{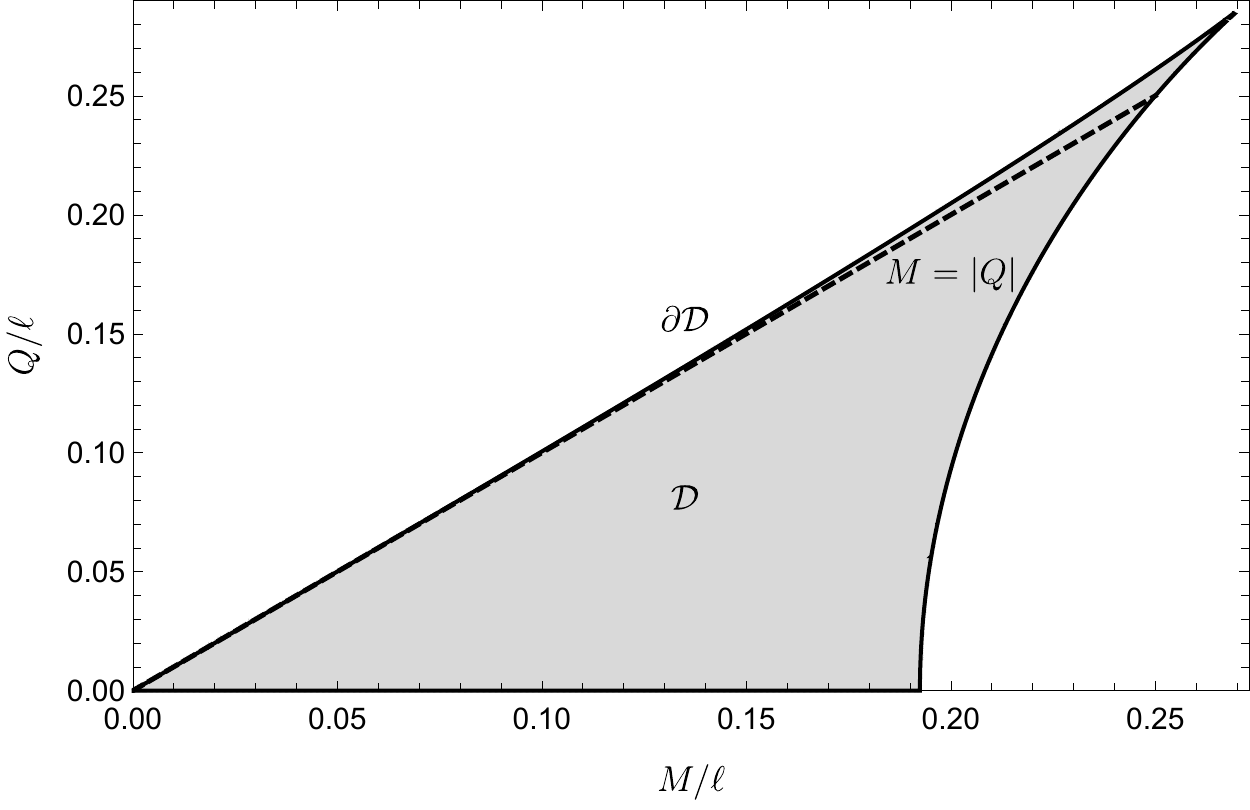}
\caption{The phase space of RNdS solutions is shown.
The gray shaded region represents the physical phase space $\mathcal{D}$.
The extremal and Nariai solutions $\partial \mathcal{D}$ are indicated by the black solid line and correspond to the boundary of the physical phase space.
The lukewarm line $M=|Q|$ is indicated with a dashed line.
The phase space outside the gray shaded region is not physical as it gives rise to naked singularities.}
\label{fig:sharkfinxy}
\end{figure}

In Fig.~\ref{fig:sharkfinxy}, the physical phase space $\mathcal{D}$ and its boundary $\partial\mathcal D$ are shown.
Since the discriminant locus \eqref{delta} and the horizons \eqref{eq:rhorizon} depend only on the dimensionless ratios $M/\ell$ and $Q/\ell$, the 3-dimensional phase space can be represented in a 2-dimensional plane.
Along the Nariai and cold lines, depicted in black, two horizons are degenerate and in thermal equilibrium at zero temperature, while the non-degenerate horizon has finite temperature.
The lukewarm line characterized by $M=|Q|$ and indicated by the dotted black line is special because the black hole and the cosmological horizon are in thermal equilibrium at a finite temperature, even though the horizons are not degenerate~\cite{Romans:1991nq}.
Finally, the neutral case ($Q=0$) corresponds to Schwarzschild-de Sitter.
The point $M/\ell=Q/\ell=0$ represents pure de Sitter, Schwarzschild and Reissner-Nordstr\o m simultanously.
For a recent study of these solutions in the context of the Weak Gravity Conjecture, we refer to \cite{Antoniadis:2020xso}.  

To finish this section, we introduce the concept of a \textit{geodesic observer}~\cite{Bousso:1996au}, which is crucial for this paper.
The geodesic observer is an observer that remains at their radial position without acceleration.
In asymptotically flat space-times, the geodesic observer is at infinity $\mathscr{I}^\pm $.
For asymptotic de Sitter, future and past infinity are at spatial distance due to the presence of the cosmological horizon.
The geodesic observer is located at the position $r_{\rm g}$, where the gravitational pull of the black hole and the expansion cancel.
Hence, $r_{\rm g}$ is determined by the equation
\begin{equation}\label{eq:definition-geo-obs}
    V'(r)|_{r=r_\geo} = 0\,.
\end{equation}
The geodesic observer is crucial to correctly normalize the Killing vector fields.
Take the time-like Killing vector field $\xi_{(t)} = \gamma_{(t)} \partial_t$ with constant $\gamma_{(t)}$.
Instead of normalizing at infinity, where $\xi_{(t)}$ is spacelike, the Killing vector field is normalized such that $\xi_{(t)}^2|_{r=r_{\rm g}} = -1$.
We thus have to choose $\gamma_{(t)} = V(r_{\rm g})^{-1/2}$.
This yields an extra redshift factor in the surface gravity,
\begin{equation}\label{eq:kappaNormalizedrg}
    \kappa_h = \frac{1}{2\sqrt{V(r_{\rm g})}} |V'(r_{\rm h})|\,,
\end{equation}
where the subscript denotes horizon, $h={\pm,{\rm c}}$.
In the limit of vanishing cosmological constant we recover the standard normalization because $r_{\rm g}\rightarrow\infty$ and thus $V(r_{\rm g})\rightarrow 1$.

\section{The moduli space of spherically symmetric and static space-times}
\label{sec:ModuliSpaceAndGeodesics}
Each point of the phase space characterizes a metric of the RNdS family.
In this paper, we are interested in measuring the distance between such metrics.
The concept of distance between metrics was first introduced by de Witt in \cite{DeWitt:1984cck} and extended by Gil-Medrano and Michor in \cite{Gil:1991} by constructing the geometric space of metrics.
Further, a prescription for computing the distance along a path in the moduli space of metrics was given in~\cite{Bonnefoy:2019nzv}\footnote{Also see~\cite{Kehagias:2019akr} for a detailed discussion on how to define distances between background fields in terms of entropy functionals.}.
Here, we briefly introduce the construction for a general moduli space of metrics and quickly specialize to spherically symmetric and static space-times.

Each point in the moduli space of metrics is a metric itself.
We can connect different metrics by a path in moduli space $\tau\mapsto g(\tau)$ where $\tau$ is an affine parameter.
For every $\tau$, the point $g(\tau)$ represents a metric.
We can assign a distance to such paths using the geometric distance formula~\cite{Bonnefoy:2019nzv}
\begin{equation}\label{Distance1}
\Delta_g = c  \int_{\tau_{\rm i}}^{\tau_{\rm f} } \left(\frac{1}{V_M} \int_M \text{vol}(g)\,
\text{tr} \left[\left(g^{-1} \frac{\partial g}{\partial \tau} \right)^2 \right] \right) ^{1/2} \dd\tau
\end{equation}
where $c\sim \mathcal{O}(1)$ is a constant, $V_M = \int_{M} \text{vol}(g)$ the volume of $M$ and $\text{vol}(g) = \sqrt{|\text{det}(g)|} \dd^n x$ the volume element.
Eq.~(\ref{Distance1}) hence describes the distance between the initial $g(\tau_{\rm i})$ and final $g(\tau_{\rm f})$ metrics along the path $g(\tau)$.
The concept of distance allows to define geodesic paths in the moduli space.
Specializing to isotropic space-times and minimizing the distance~\eqref{Distance1} leads to the geodesic equation
\begin{equation}\label{eq:geodesicEq}
     \ddot g - \dot g g^{-1} \dot g = 0\,.
\end{equation}
A dot indicates differentiation with respect to the proper time $\lambda$, which is related to the affine parameter $\tau$ via
\begin{equation}\label{propertime}
\dd\lambda = \dd\tau \left(\frac{1}{V_M}\int_M \sqrt{g}\, \text{tr}\left[\left(g^{-1}\frac{\partial g}{\partial \tau}\right)^2\right]\right)^{1/2}\,.
\end{equation}

The distance functional \eqref{Distance1} is not covariant under diffeomorphisms\footnote{See Appendix A of~\cite{Bonnefoy:2019nzv} for a detailed discussion on this issue and~\cite{Kehagias:2019akr} for a diffeomorphism invariant definition of the distance functional based on the Ricci flow. } along the metric flow\footnote{The
geodesic flow of a smooth Riemannian manifold is defined as follows~\cite{Carmo:722464}: Let $t\mapsto \gamma(t)$ be a geodesic differentiable path on $M$ and let $TM$ denote the tangent bundle of $M$.  Hence, a geodesic curve on $M$ determines a curve $t\mapsto (\gamma(t),\frac{\dd \gamma}{\dd t}(t))$ in $TM$. The flow of the unique vector filed $G$ on $TM$ whose trajectories are of the form $t\mapsto (\gamma(t),\gamma'(t))$ is called  the geodesic flow on $TM$.}~\cite{Bonnefoy:2019nzv}.
Consequently, the metric path $g(\lambda)$ along which the distance is computed needs to be expressed in a specific coordinate system. 
It is convenient to work in Eddington-Finkelstein gauge with dimensionless coordinates $(\tilde v,\tilde r, \theta,\phi)$, where $(v,r) = \alpha(\tilde v, \tilde r)$.
Here $\alpha$ is a parameter of mass dimension one.
Updating $\alpha\rightarrow \alpha(\lambda)$, the metric flow $g(\lambda)$ is entirely captured by the flow of $\alpha(\lambda)$, while the dimensionless coordinates $(\tilde v, \tilde r)$ are constant along the flow.
In this gauge the line element~\eqref{metricRNdS} reads
\begin{equation}\label{eq:dsEFalpha}
\text{ds}^2 =-\alpha^2V(\alpha \tilde r,\alpha_i) \text{d}\tilde v^2 + 2 \alpha^2 \text{d}\tilde v \text{d} \tilde r + \alpha^2 \tilde r^2 \text{dS}_2^2\,.
\end{equation}
The metric potential $V(\alpha \tilde r, \alpha_i)$ might depend on further parameters $\alpha_i(\lambda)$ of mass dimension one, that also flow.

Using the above prescription, the geodesic equation~\eqref{eq:geodesicEq} in dimensionless Eddington-Finkelstein gauge reduces to the following set of equations,
\begin{align}\label{eq:conditions}
\frac{\dd}{\dd\lambda} \left(\frac{\dot{\alpha}}{\alpha}\right) = 0\,,\quad
\frac{\dd^2}{\dd\lambda^2}V(\alpha \tilde r, \alpha_i) = 0\,.
\end{align}
The first equation is solved by
\begin{equation}\label{eq:sol-alpha}
    \alpha(\lambda) = \alpha_{\rm i} e^{\lambda/4}\,,
\end{equation}
where $\alpha_{\rm i}$ is a constant of integration.
In section~\ref{sec:Evolution_Mass_Parameters_Moduli} we will provide a general solution to the second equation.
However, we can already relate the geometric distance to the proper time $\lambda$.
Noting that ${\rm tr}(\dot g g^{-1}g g^{-1}) \propto (\dot\alpha/\alpha)^4$, the geometric distance is given by
\begin{equation}\label{eq:distance-final}
    \Delta = 4c \log\left(\frac{\alpha(\lambda_{\rm f})}{\alpha(\lambda_{\rm i})}\right) = c |\lambda_{\rm f} - \lambda_{\rm i}|
\end{equation}
upon using Eq.~\eqref{eq:sol-alpha}.
Hence, we can already conclude that the geometric distance between metrics is infinite, if they are separated by infinite moduli space proper time.
This simple relation was previously established only for Weyl rescalings~\cite{Lust:2019zwm,Bonnefoy:2019nzv}.

The distance conjecture relies on the moduli space of fields having negative curvature~\cite{Ooguri:2006in}.
In Appendix~\ref{sec:AppB} we explicitly show that the moduli space of spherically symmetric and static space-times indeed has negative scalar curvature.

%%%%%%%%%%%%%%%%%%%%%%%%%%%%%%%%%%%%%%%%%%%%%%%%%%%%%%%%
\subsection{Interlude: General solution via exponential mapping}
\label{sec:exponentialMap}
%%%%%%%%%%%%%%%%%%%%%%%%%%%%%%%%%%%%%%%%%%%%%%%%%%%%%%%%

In this section we provide a general solution to the geodesic equation~\eqref{eq:geodesicEq} for isotropic space-times using exponential mapping. 
Our result demonstrates that dimensionless Eddington-Finkelstein (EF) gauge is indeed the most convenient choice for computing geometric distances. From this point onward we suppress the explicit dependence of the map $g(\lambda)(\tilde x)$ on the dimensionless space-time coordinates for simplicity of notation, since the dimensionless space-time coordinates are fixed along the flow. 

Let $g: [0,1]\times M \to \mathcal{M}$ be a smooth geodesic in moduli space. 
The evolution of $g(\tau)$ only depends on the initial point $g(0)$ and the tangent vector $\dot{g}(0)$ at the initial point for fixed space-time coordinates.
Indeed, the moduli space evolution is such that\footnote{Eq.~\eqref{eq:geodesicEq} is an ordinary differential equation (ODE).
By the existence and uniqueness theorems for ODEs, the solution of the geodesic equation exists and is unique in an open neighbourhood of a smooth manifold, once a point and a tangent vector are specified~\cite{Carmo:722464}.
}
$g(\lambda) \in \mathcal{M}$, $\dot{g}(\lambda)\in T_{g(\lambda)}\mathcal{M}$ $\forall \lambda\in [0,1]$~\cite{Gil:1991}.
With $g_{\rm i} = g(\lambda=0)$, the geodesic in $\mathcal{M}$ starting at $g_{\rm i}$ in the direction $\dot{g}(0)$ is given by
\begin{equation}\label{eq:geodesicsolgeneral}
    g(\tau) = g_{\rm i} \exp\Big[a(\lambda)\text{Id} + b(\lambda)H_0\Big]
\end{equation}
where $a(\lambda) = c_a \lambda$ and $b(\lambda) = c_b\lambda$ are smooth functions with some real constants $c_a$ and $c_b$.
Further, $\text{Id}$ is the identity matrix.
We have introduced the quantity $H = g_{\rm i}^{-1} \dot{g}(0)$ and its traceless part $H_0 = H-n^{-1}\text{tr}(H) \text{Id}$
with $n$ the dimension of the space-time $M$.
We specify to 4-dimensional Lorentzian manifolds, $n=4$.
The specific values of $c_a$ and $c_b$ depend on the initial point $g_{\rm i}\in \mathcal{M}$ and the initial direction, and hence on the dimensionless space-time coordinates.

The general solution (\ref{eq:geodesicsolgeneral}) allows to compute the geometric distance in any gauge.
In dimensionless Eddington-Finkelstein coordinates~\eqref{eq:dsEFalpha}, we have $c_a = 2\dot\alpha/\alpha$ and $c_b =\dot{V}(\lambda)/V(0)$. Here, $V(\lambda)$ denotes the metric potential \eqref{eq:V_norm} that might depend on $\lambda$ via the mass parameters and on the dimensionless space-time coordinates. The constant $c_a$ is independent of the space-time coordinates since we consider isotropic space-times. 
This yields for the geodesic path
\begin{equation}\label{eq:geodesicsolution}
g(\lambda) = g_{\rm i} \exp \Bigg[2 \log \left(\frac{\alpha(\tau)}{\alpha(0)}\right) \text{Id} + \frac{V(\tau)-V(0)}{\dot{V}(0)}\,H_0\Bigg]\,.
\end{equation}
This path $g(\lambda)$ smoothly connects $g_{\rm i} = g(\lambda_{\rm i})$ and $g_{\rm f} = g(\lambda_{\rm f})$.
In this gauge we find that $\text{tr}H_0^2=0$.
Hence, computing the geodesic distance along this path is remarkably simple.
The result is provided by Eq.~\eqref{eq:distance-final}.

This is not the case in other gauges.
As an example, we work out the case of dimensionless Schwarzschild gauge.
We introduce dimensionless coordinates via $(t,r) = \alpha (\tilde t,\tilde r)$.
The general solution to the geodesic equation is then given by
\begin{equation}
g(\tau) = \tilde g_{\rm i} \exp\Bigg[ 2 \log \left(\frac{\alpha(\lambda)}{\alpha(0)}\right) \text{Id} + \frac{V(0)}{\dot{V}(0)} \log\left(\frac{V(\tau)}{V(0)}\right) H_0\Bigg]\,.
\end{equation}
It turns out that the second term does not vanish when evaluating the trace in the distance formula~\eqref{Distance1} because $\text{tr} H_0^2\ne0$.
Since the term containing $H_0$ encodes how the geodesic path explicitly depends on space-time coordinates $\tilde x$, we expect the distance computed in such gauges to not correspond to the distance between space-times.
However, it would be interesting to explicitly evaluate the integral in the distance formula to check whether the term proportional to ${\rm tr} H_0^2$ evaluates to zero.

%%%%%%%%%%%%%%%%%%%%%%%%%%%%%%%%%%%%%%%%%%%%%%%%%%%%%%%%
\subsection{Flow of mass parameters}
\label{sec:Evolution_Mass_Parameters_Moduli}
%%%%%%%%%%%%%%%%%%%%%%%%%%%%%%%%%%%%%%%%%%%%%%%%%%%%%%%%

So far, we have solved only parts of the geodesic equation.
In this section, we solve the second equation of~\eqref{eq:conditions}.
For RNdS metrics, the mass parameters $\alpha_i$ that flow with $\lambda$ are the mass $M$, charge $Q$ and de Sitter radius $\ell$.
Hence, we promote them to depend on $\lambda$.

The second differential equation of~\eqref{eq:conditions} can be separated into three independent differential equations by requiring isotropy.
In other words, the functions $M(\lambda)$, $Q(\lambda)$ and $\ell(\lambda)$ must be independent of $\tilde r$.
Using the general solution for $\alpha$~\eqref{eq:sol-alpha}, the resulting differential equations are given by
\begin{align}\label{eq:diffEqgeoSol}
\begin{split}
 \alpha_{\rm i}^2 M - 2 \alpha_{\rm i} \dot M + \ddot M &=0\,,\\
  2 \alpha_{\rm i}^2 Q^2 + \dot Q^2 + Q(-4 \alpha_{\rm i} \dot Q + \ddot Q ) &=0\,,\\
 2 \alpha_{\rm i}^2 \ell^2 + 3 \dot\ell^2 -\ell(4 \alpha_{\rm i}\dot\ell+\ddot\ell) &=0\,.
\end{split}
\end{align}
We have thus decoupled the system of differential equations and can solve them separately for each mass parameter.
The solutions read
\begin{align}\label{eq:mqloflambda}
\begin{split}
M(\lambda) &=  M_{\rm i} e^{\lambda/4}(1+c_M \lambda)\,,\\
Q(\lambda) & =Q_{\rm i} e^{\lambda/4}\sqrt{1+c_Q \lambda}\,,\\
\ell(\lambda) & = \frac{\ell_{\rm i} e^{\lambda/4}}{\sqrt{1+c_\ell \lambda}}\,,
\end{split}
\end{align}
where we have chosen the integration constants such that $M_{\rm i} = M(\lambda=0)$ and likewise for the other parameters w.l.o.g.
Further, also $c_M$, $c_Q$ and $c_\ell$ are integration constants.
These general solutions are the basis for studying the geometric distance between space-times that belong to the RNdS as done in the next section.

%%%%%%%%%%%%%%%%%%%%%%%%%%%%%%%%%%%%%%%%%%%%%%%%%%%%%%%%
\subsection{Distances between specific space-times}
\label{sec:Distance}
%%%%%%%%%%%%%%%%%%%%%%%%%%%%%%%%%%%%%%%%%%%%%%%%%%%%%%%%

In this section, we will systematically compute the distance between different space-times that belong to the RNdS family.
We are particularly interested in the distance to Minkowski space-time, but we will also study the distances between other points in the phase space $\mathcal D$.

We found the general solutions for $\alpha$ and the mass paramaters $M$, $Q$ and $\ell$ as functions of proper time $\lambda$.
However, $\alpha$ is only a parametrization to study the metric flow, but does not correspond to a mass parameter.
Instead, since $\alpha$ must be a smooth function of the mass parameters, they have to be related in a way that is compatible with Eqs.~\eqref{eq:sol-alpha} and \eqref{eq:mqloflambda}.
We use a power-law ansatz,
\begin{align}
\begin{split}
\label{eq:alphaConstantsExponents}
    \alpha(\lambda)
    &\propto \sum_{\beta,\gamma,\delta} M(\lambda)^\beta Q(\lambda)^\gamma \ell(\lambda)^\delta\\
    &=\sum_{\beta,\gamma,\delta} M_{\rm i}^\beta Q_{\rm i}^\gamma \ell_{\rm i}^\delta \frac{(1+c_M \lambda)^\beta(1+c_Q\lambda)^{\gamma/2}}{(1+c_\ell \lambda)^{\delta/2}} e^{\lambda(\beta+ \gamma+\delta)/4}\,, \end{split}
\end{align}
where the sum is subject to the condition $\beta + \gamma+\delta = 1$ due to dimensional reasons.
We immediately see that Eq.~\eqref{eq:sol-alpha} implies the conditions $2\beta+\gamma-\delta = 0$ and $c_M=c_Q=c_\ell$ if all exponents are non-zero.
Then we can also express the inital value of $\alpha$ in terms of the mass parameters as $\alpha_{\rm i} = M_{\rm i}^\beta Q_{\rm i}^\gamma \ell_{\rm i}^\delta$.

Specializing to $c_M=c_Q=c_\ell=0$, our general solutions reduce to Weyl rescalings because the metric function~\eqref{eq:V_norm} $V$ is independent of $\lambda$ in that case.
Any choice of exponents respecting the condition $\beta+\gamma+\delta=1$ constitutes a valid parametrization.
We hence recover the framework of~\cite{Lust:2019zwm,Bonnefoy:2019nzv}.
As an example, we consider the RN black hole.
Since the metric function does not depend on $\ell$ we set $\delta=0$.
The choice $\gamma=0$ leads to $\alpha(\lambda)= M(\lambda)$ so the metric flow is characterized by~\cite{Bonnefoy:2019nzv}
\begin{equation}
    M(\lambda) = M_{\rm i} e^{\lambda/4}\,, \quad Q(\lambda) = Q_{\rm i} \sqrt{ 1+c_Q \lambda}\,.
\end{equation}
The other cases discussed in~\cite{Bonnefoy:2019nzv} are reproduced by similar choices for $\beta$, $\gamma$ and $\delta$.

In the following, we will use different combinations of $\beta$, $\gamma$ and $\delta$ and $c_M$, $c_Q$ and $c_\ell$ to describe geodesics that connect specific space-times to infer the geodesic distance between them.

%%%%%%%%%%%%%%%%%%%%%%%%%%%%%%%%%%%%%%%%%%%%%%%%%%%%%%%%
\subsubsection{The distance to Minkowski space-time}
\label{sec:distance-minkowski}
%%%%%%%%%%%%%%%%%%%%%%%%%%%%%%%%%%%%%%%%%%%%%%%%%%%%%%%%

From the perspective of RNdS space-time, there are two distinct Minkowski limits.
The first and trivial limit is characterized by vanishing mass parameters $M=Q=\ell^{-1}=0$.
We will denote this limit as Mink${}_0$ in the following.
The second Minkowski limit is arrived at by taking the black hole mass to infinity, $M\rightarrow\infty$.
To avoid naked singularities, also the other mass parameters have to scale as $Q\rightarrow\infty$ and $\ell\rightarrow\infty$.
As demonstrated in~\cite{Averin:2016ybl,Bonnefoy:2019nzv}, the near horizon geometry approximates that of Minkowski in this limit.
Further, the $BMS$ groups at both the black hole and cosmological horizons approach those of future and past null infinity.
We will denote this Minkowski limit as Mink${}_\infty$ in the following.
In this section we infer the geometric distance of any space-time in the RNdS family to both ${\rm Mink}_{0,\infty}$.

First, we study the distance to Mink${}_\infty$.
We show that this limit is at infinite distance from any spherically symmetric and static space-time independently of the parametrization $\alpha$.
From Eq.~\eqref{eq:mqloflambda} we immediately see that the infinite mass and charge limit is reached only for $\lambda\to\infty$.
Hence, Eq.~\eqref{eq:distance-final} implies that the distance to Mink${}_\infty$ is always infinite,
\begin{equation}
    \Delta = c|\lambda_{\rm f}-\lambda_{\rm i}|\to \infty\,.
\end{equation}
Our result generalizes the discussion about the infinite distance limits of~\cite{Bonnefoy:2019nzv} to arbitrary spherically symmetric and static space-times.
Notice that the proof above  was independent of the parametrization $\alpha$, and hence of the specific moduli space geodesic, and of the particular causal structure of the initial space-time.
Hence, Mink${}_\infty$ is a moduli space point that represents Minkowski space-time and lies at an infinite distance of any other space-time independently of the geodesic path used to reach it. 

We continue by discussing the distance to Mink${}_0$.
Naively, this limit can be reached in two ways.
The first option to arrive at vanishing mass $M$ is $\lambda\to -\infty$.
To avoid naked singularities in this limit we have to impose $c_M<0$, $c_Q<0$ and $c_\ell<0$ according Eq.~\eqref{eq:mqloflambda}.
However, in this limit not only $M\to 0$ and $Q\to 0$, but also $\ell \to 0$ implying that the scalar curvature diverges, $R = 12/\ell^2\to \infty$.
Hence, this is not a correct Minkowski limit.
The second option to achieve a vanishing mass $M$ is given by $\lambda \to -1/c_M$.
However, $Q\to 0$ and $\ell\to\infty$ need to be reached simultaneously.
This enforces the integration constants to satisfy $c_M = c_Q=c_\ell \equiv k$.
Then, Mink${}_0$ is reached at proper time $\lambda = -1/k$.
Notice that the conditions $M\geq0$ and $Q,\ell \in \mathrm{R}$ together with the absence of naked singularities at the flow start and endpoints\footnote{This condition can be imposed simply by requiring that the discriminant locus~\eqref{delta} of the metric function is positive $D\geq0$.} restricts the range of the proper time completely to a bounded interval.
Setting $k=-1$ w.l.o.g. yields the proper time to lie in the interval $\lambda\in [0,1]$.
This means that the geodesics to Mink${}_0$ are unique because all arbitrary constants in Eq.~\eqref{eq:mqloflambda} are fixed and the parametrization $\alpha(\lambda)$ is fully specified by Eq.~\eqref{eq:alphaConstantsExponents}.
We proceed by explicitly constructing minimal geodesic paths connecting RNdS space-times as well as its various subcases to Mink${}_0$.

Let us consider SdS space-time.
This case does not contain $Q$ so we set $\gamma=0$.
Hence, Eq.~\eqref{eq:alphaConstantsExponents} simplifies to
\begin{equation}
    \alpha\propto M^{1/3} \ell^{2/3}\,. 
\end{equation}
This coincides with the expression for the position of the geodesic observer in SdS space-time,
\begin{equation}\label{eq:rgeoSdS}
     r_\geo = (M \ell^2)^{1/3} = (M_{\rm i} \ell_{\rm i}^2)^{1/3} e^{\lambda/4}\,,
\end{equation}
as follows from Eq.~\eqref{eq:definition-geo-obs}.
The parametrization is a smooth function of the mass parameters, finite, real-valued and strictly positive for all values of $\lambda\in [0,1]$.
Thus, the geodesic path is characterized by the parametrization $\alpha= r_\geo$ and the mass parameter evolution follows Eq.~\eqref{eq:mqloflambda} with $c_{M,Q,\ell} = -1$ and $\lambda\in [0,1]$. 
We move on to the RN black hole, for which we set $\delta = 0$ because this case does not contain $\ell$.
Hence we obtain the parametrization
\begin{equation}
    \alpha = \frac{Q^2}{M} = \frac{Q_{\rm i}^2}{M_{\rm i}} e^{\lambda/4} = r_{\geo}\,,
\end{equation}
which again coincides with the expression for the geodesic observer for the RN space-time\footnote{The 
RN space-time presents two geodesic observers.
One is located at infinity because the space-time is asymptotically flat.
The other one is located between the Cauchy and the black hole horizon, given by $ r_\geo = Q^2/M$.}.
Finally, the same applies for the RNdS space-time.
The parametrization $\alpha=r_\geo = r_{\geo,\rm i} e^{\lambda/4}$ defines a smooth path connecting a non-Nariai RNdS space-time to Mink${}_0$.
Since the expression is lengthy, we do not display it here explicitly.

We conclude that the path $\lambda \mapsto g(\lambda)$ parametrized by $\alpha = r_\geo$ is positive and smoothly connects SdS, RN and RNdS space-times to Mink${_0}$.
The corresponding distances are given by 
\begin{equation}\label{eq:Deltag}
    \Delta_\geo = 4c \log\left(\frac{r_{\geo,\text{f} }}{ r_{\geo, \text{i}}}\right) = 4c\,,
\end{equation}
where we used that $\lambda_{\rm i}=0$ and $\lambda_{\rm f}=1$.
The distance to Mink${}_0$ is hence always finite.

The discussion above applies to space-times with more than one mass parameter, i.e. SdS, RN and RNdS, for which the geodesic observer is located at a finite radial position.
We finish by analyzing the geodesics connecting S and dS to Mink$_0$. 
If we set the constants to $c_M=0$ or $c_\ell=0$, the geodesics to Mink${}_0$ are parametrized by $\alpha\propto M$ or $\alpha\propto\ell$, respectively, i.e. Weyl rescalings.
Hence, there are geodesics connecting both Mink${}_\infty$ and Mink${}_0$ to Schwarzschild space-time at infinite distance, since these limits are obtained as
 \begin{equation}
     \lim_{\lambda\to \infty} M=\infty\,,\quad \lim_{\lambda\to -\infty} M = 0\,.
 \end{equation}
For dS, Mink$_0$ is arrived in the limit $\lim_{\lambda\to \infty} \ell = \infty$.
Hence, the limit Mink${}_0$ is at infinite distance from both S and dS.
If we instead allow for $c_M\neq0$ or $c_\ell\neq 0$, respectively, geodesics exist that connect both S and dS to Mink$_{0}$ at finite distance.
Since the horizons are not co-moving with the geometry, we need to impose extra conditions on the parameters to ensure consistency along the flow.
For the case of S, the mass of the black hole has to be positive and any observer outside the black hole must not cross the black hole horizon.
These conditions yield
\begin{subequations}
\begin{align}
    M\geq 0\quad &\Longrightarrow\quad c_M \lambda \geq -1\,,\\
    r\geq 2M\quad &\Longrightarrow\quad c_M \lambda \leq \frac{r_{\rm i}}{2M_{\rm i}} -1 \,,
\end{align}
\end{subequations}
and imply that the range of proper time must lie in the finite range $\lambda \in [-1/c_M,(r_{\rm i}/(2M_{\rm i})-1)/c_M]$.
Setting $c_M=-1$ and specifying to an observer that is initially located on the black hole horizon, $r_{\rm i}=2M_{\rm i}$, the range of $\lambda$ is $\lambda\in [0,1]$.
The mass and radius evolve as
\begin{equation}
    M(\lambda) = M_{\rm i} e^{\lambda/4} (1-\lambda)\,,\quad r(\lambda) = r_{\rm i} e^{\lambda/4} \,,
\end{equation}
so the black hole shrinks to zero while the observer drifts away.
This parametrization yields the distance between S and Mink$_0$ as
\begin{equation}\label{eq:DistanceSanddS}
    \Delta = 4c \log \left(\frac{r_{\rm f}}{r_{\rm i}}\right) \,.
\end{equation}
However, in this case the horizons are not co-moving with the radial coordinates.
That means that this notion of distance receives an extra contribution, which does not stem from the pure distance between space-times.
Hence, Eq.~\eqref{eq:DistanceSanddS} gives an upper bound on the distance.
The discussion for de Sitter space-time is completely analogous\footnote{For dS impose that the observer must not cross the cosmological horizon and that the de Sitter radius is real-valued.}.
We conclude that the distance between the S and dS to Mink${}_0$ is finite and bounded from above by~\eqref{eq:DistanceSanddS}.

To summarize, Mink${}_\infty$ is at infinity distance from RNdS and all its subcases.
On the other hand, parametrizing the metric flow in terms of the geodesic observer, $\alpha=r_\geo$, allows to construct geodesics connecting RN, dS and RNdS to Mink${}_0$.
This establishes that the distance between these space-times is finite.
For S and dS, the flow cannot be parametrized in terms of the geodesic observer, but we find that these space-times are at finite distance from Mink${}_0$ as well.
In the special case of Weyl-rescalings the distance to Mink${}_0$ turns out to be infinite in all cases.

\subsubsection{Distances between space-times with non-zero mass parameters}
\label{sec:Distances_between_space-times_withnonzeromass}

Eq.~\eqref{eq:alphaConstantsExponents} allows to identify different geodesics connecting spherically symmetric and static space-times with non-vanishing mass parameters.
These geodesics are not unique, since the choice of initial and final space-times only fixes some of the integration constants $c_M$, $c_Q$ and $c_\ell$, but not all of them.
Hence, different combinations of exponents and constant yield different geodesics. 
A simple strategy for identifying geodesic is the following: set one of the relevant constants appearing on Eq.~\eqref{eq:alphaConstantsExponents} to zero and study the allowed range of $\lambda$ dictated by the remaining mass parameters. Here, we present some relevant examples of geodesics between space-times. 

We start by analyzing the geodesic paths from RNdS, SdS and RN to S.
The geodesics describing the flow to the Schwarzschild geometry must be such that the mass of the black hole remains finite while $Q\to 0$ and $\ell \to \infty$ smoothly.
A simple choice is to set $c_M=0$ implying that $\alpha\propto M$ as then the mass does not vanish for finite values of $\lambda$.
In the following we identify geodesics connecting these space-times to S and infer the geometric distance.
\begin{itemize}
\item
RN $\to$ S:
Let's start by setting $c_M=0$.
Then, $M=M_{\rm i} e^{\lambda/4}\propto \alpha$ and $Q = Q_{\rm i} e^{\lambda/4} \sqrt{1+c_Q \lambda}$.
The charge of the black hole needs to be real-valued and the black hole must remain subextremal along the flow of mass parameters.
Hence, 
\begin{subequations}
\begin{align}
    Q\in \mathbf{R}&\quad \Longrightarrow\quad \lambda\geq -1/c_Q\,, \\
    M\geq Q &\quad \Longrightarrow\quad \lambda\leq \frac{1}{c_Q} \left(M_{\rm i}^2/Q_{\rm i}^2 -1\right)\,,
\end{align}
\end{subequations}
so the range of proper time is bounded $\lambda\in [-1/c_Q,((M_{\rm i}/Q_{\rm i})^2-1)/c_Q]$.
This geodesic describes the flow of an extremal RN black hole which loses all its charge until it becomes a Schwarzschild black hole.
The distance is finite and given by
\begin{equation}
    \Delta = 4 c \log \left(\frac{ r_{+,\rm f}}{r_{+,\rm i} + r_{-,\rm i} }\right)\,,
\end{equation}
where $r_{+,\rm f} = 2 M_{\rm f}$ and $r_\pm = M\pm \sqrt{M^2-Q^2}$.
This follows from $\alpha = M = (r_+ + r_-)/2$.
Alternatively,  the geodesic equation can also be solved for the position of the horizons
\begin{equation} \label{eq:SolRNSubextremalNew}
     r_\pm = r_{\pm,\rm i} e^{\lambda/4}\,,\quad r_\mp = r_{\mp,\rm i} e^{\lambda/4} (1+c_\mp \lambda)\,.
\end{equation}
Choosing $r_+ = r_{+,\rm i} e^{\lambda/4}$ enforces the consistency conditions\footnote{This solution is equivalent to~\eqref{eq:mqloflambda} by identifying $c_M = c_Q r_{(-,\rm i)}/(2M_{\rm i})$ and $c_-=c_Q$. } 
\begin{subequations}
\begin{align}
    r_-\geq 0 & \quad \Longrightarrow\quad \lambda \geq -\frac{1}{c_-}\,,\\
    r_-\leq r_+ & \quad\Longrightarrow\quad \lambda\leq \frac{1}{c_-} \left(\frac{r_{+,\rm i}}{r_{-,\rm i}}-1\right)\,.
\end{align}
\end{subequations}
Hence, the solution of the geodesic equation~\eqref{eq:SolRNSubextremalNew} also describes a geodesic connecting an extremal or subextremal RN black hole to S.
The distance is finite and explicitly given by
\begin{equation}
    \Delta = 4c \log \left(\frac{r_{+,\rm f}}{r_{+,\rm i}}\right) \,. 
\end{equation}
Notice that the first geodesic is valid for an observer between the interior and exterior black hole horizons, while the second one considers an observer outside of the exterior black hole horizon.

\item
SdS $\to$ S:
A simple choice is $c_M=0$ such that $\lambda\in [0,-1/c_\ell]$.
Then,
\begin{equation}
    \alpha\propto M = \frac{r_+ r_\cc (r_+ + r_\cc)}{2(r_+^2 + r_\cc^2 + r_+ r_\cc)}\,.
\end{equation}
The limit $\ell\to \infty$ yields $\alpha\propto r_{+}$.
If we instead set $c_M=0.17$ and $\ell=-1$ we find that $r_+=r_{+,\rm i}e^{\lambda/4}$ such that the geodesic is parametrized by $\alpha\propto r_+$ as we have checked numerically.

\item
RNdS $\to $ S:
There exist a geodesic path connecting the RNdS space-time to Schwarzschild characterized by the constants $c_Q=c_\ell=k$, such that the charge and cosmological constant vanish at $\lambda=-1/k$.
Then, setting $c_M=0$ implies $\alpha \propto M$,
\begin{equation}
    \alpha =  \frac{(r_{+} +r_\cc)(r_++r_-)(r_\cc+r_-)}{2(r_+^2 + r_\cc^2 + r_-^2+ r_-r_\cc +r_-r_++r_\cc r_+)}
\end{equation}
which reduces to $\alpha(\lambda=-1/k) = r_{+,\rm f}=2M_{\rm f}$.
Alternatively, upon setting $c_Q=c_\ell$ we can choose
\begin{equation}
    \alpha = \sqrt{Q \ell}=  \left((r_+r_-r_\cc(r_++r_-+r_\cc)\right)^{1/4}\,,
\end{equation}
which also reduces to $\alpha(\lambda=-1/k) = r_{+,\rm f}$.

\end{itemize}

Next, we consider geodesic paths connecting RNdS to RN.
These geodesics describe the divergence of the de Sitter radius while the mass and charge of the black hole remain finite.
It suffices to set $c_\ell=k$ and consider the proper time interval $\lambda\in [0,-1/k]$.
There are different possibilities for the constants $c_M$ and $c_Q$, so that the geodesic yields the RN space-time.
One such possibility is to set $c_M=c_Q=0$ and consider the parametrization 
\begin{align}
\begin{split}
    \alpha &= \frac{(r_++r_\cc)(r_++r_-)(r_\cc+r_-)}{2\ell^2} \\
   & +\sqrt{\frac{(r_++r_\cc)^2(r_++r_-)^2(r_\cc+r_-)^2}{4\ell^4}-\frac{r_+r_-r_\cc(r_-+r_++r_\cc)}{\ell^2}}\propto e^{\lambda/4}\,. 
\end{split}
\end{align}
Then, in the limit $\lambda\to -1/k$, we recover $\alpha\to r_+$ with $r_+ = M+\sqrt{M^2-Q^2}$ the  exterior horizon of the RN black hole. 

Further, the geodesic paths leading to dS are also finite.
The dS space-time is reached as the zero mass (and charge) limit of the SdS (or RNdS) geometries.
We consider both cases separately:
\begin{itemize}
\item
SdS $\to$ dS:
The mass of the black hole needs to vanish at the endpoint of the metric flow.
Hence, setting $c_\ell = 0$ and $c_M\neq 0$ yields
 $\ell = \ell_i e^{\lambda/4}$ and $M =M_{\rm i} e^{\lambda/4} (1+c_M \lambda)$. The consistency conditions imposed by the CCC and the positivity of the black hole mass restrict the proper time to a finite interval
\begin{subequations}
\begin{align}
    \ell \geq 3\sqrt{3} M & \quad \Longrightarrow\quad \lambda \leq \frac{1}{c_M} \left(\frac{1}{\sqrt{27}}\frac{\ell_{\rm i}}{ M_{\rm i}}-1\right) \,,\\
    M\geq 0 &\quad\Longrightarrow\quad \lambda \geq -1/c_M\,.
\end{align}
\end{subequations}
The geodesic characterized by this flow of the mass parameters connects dS to SdS.
Then, the parametrization $\alpha$ can be chosen to be $\alpha\propto \ell = \sqrt{r_\cc^2+r_+^2+r_+ r_\cc}$, so the distance between SdS and dS is given by 
\begin{equation}
    \Delta =4 c \log\left(r_{\cc,\rm f}\frac{1}{r_{+,\rm i}^2+r_{\cc,\rm i}^2+r_{+,\rm i}r_{\cc,\rm i}}\right)\,.
\end{equation}
The following choice of constants $c_M=-1$ and $c_\ell=0.81$.
leads to a parametrization proportional to the black hole horizon $\alpha\propto r_+$. 

\item
RNdS $\to$ dS:
Here we provide two different examples of geodesics connecting the RNdS to dS.
First, the constants $c_{M,Q,\ell}$ can be set such that $\alpha\propto r_\cc$. 
Here we state that the distance is finite and given by
\begin{equation}
    \Delta = 4c\log \left(\frac{r_{\cc,\rm f}}{r_{\cc,\rm i}}\right)\,,
\end{equation}
with $r_{\cc,\rm f} = \ell_{\rm f}$.
Alternatively, choosing $c_M=c_Q=k$ and $c_\ell=0$ yields dS as $M,Q \to 0$ and $\ell \neq 0$.
This choice of constants implies $\alpha = \ell = \sqrt{r_-^2+r_+^2+r_\cc^2+r_- r_+ + r_- r_\cc+r_+ r_\cc}$.
The distance is given by 
\begin{equation}
    \Delta =4 c\log \left(\frac{\ell_{\rm f}}{\sqrt{r_{-,\rm i}^2+r_{+,\rm i}^2+r_{\cc,\rm i}^2+r_{-,\rm i} r_{+,\rm i} + r_{-,\rm i} r_{\cc,\rm i}+r_{+,\rm i} r_{\rm,\rm i}}}\right)\,.
\end{equation}

\end{itemize}
Finally, it is left to analyze the geodesic connecting the RNdS to SdS.
Again, there exist different combinations of the mass parameters that yield the desired behaviour for $\alpha$.
The flow from RNdS to SdS is characterized by the discharge of the black hole.
Hence, $\lambda\in [0,-1/c_Q]$ with $c_Q\ne 0$.
A simple choice for the parametrization characterizing this geodesic is $c_M=0$, $c_\ell\neq c_Q$ such that $\alpha=M= (r_++r_-)(r_\cc+r_-)(r_\cc+r_+)/(2\ell^2)$.
Alternatively, setting $c_\ell=0$ and $c_M\neq c_Q$ yields $\alpha=\ell = \sqrt{r_-^2+r_+^2+r_\cc^2+r_- r_+ + r_- r_\cc+r_+ r_\cc}$.
Furthermore, the choice $c_M=c_\ell$ yields the parametrization $\alpha=(M \ell^2)^{1/3}$.

\subsubsection{Summary}

We summarize the results of this section in Fig.~\ref{fig:Conclusion_scheme}, where the distances between space-times are schematically represented.
First we demonstrated that Mink${}_\infty$ is at infinite distance from all the space-time configurations that fall into the RNdS class.
This is represented by the red dashed lines.
On the other hand, Mink${}_0$ is at finite distance from all space-time configuration belonging to the RNdS class.
This is indicated by the black solid lines.
Note the ambiguity in inferring the distance from S and dS to Mink${}_0$ discussed previously.

Going further, we studied geodesics and the corresponding geometric distances between various space-time configurations with different number of mass parameters that belong to the RNdS family.
There always exist a family of geodesics connecting different space-time configuration characterized by the free constants of integration.
However, specific choices are more convenient as we have demonstrated case-by-case.
In more detail, to describe the flow between SdS, RN and RNdS the parametrization in terms of the geodesic observer, $\alpha=r_\geo$ is particularly convenient.
As a result, we find that all space-time configurations characterized by at least one mass parameter are at finite distance from each other.
This is indicated by the green lines in Fig.~\ref{fig:Conclusion_scheme}.

\tikzstyle{block} = [rectangle, draw, fill=blue!30!, 
    text width=5em, text centered, rounded corners, minimum height=4em]
\tikzstyle{line} = [draw, -latex']
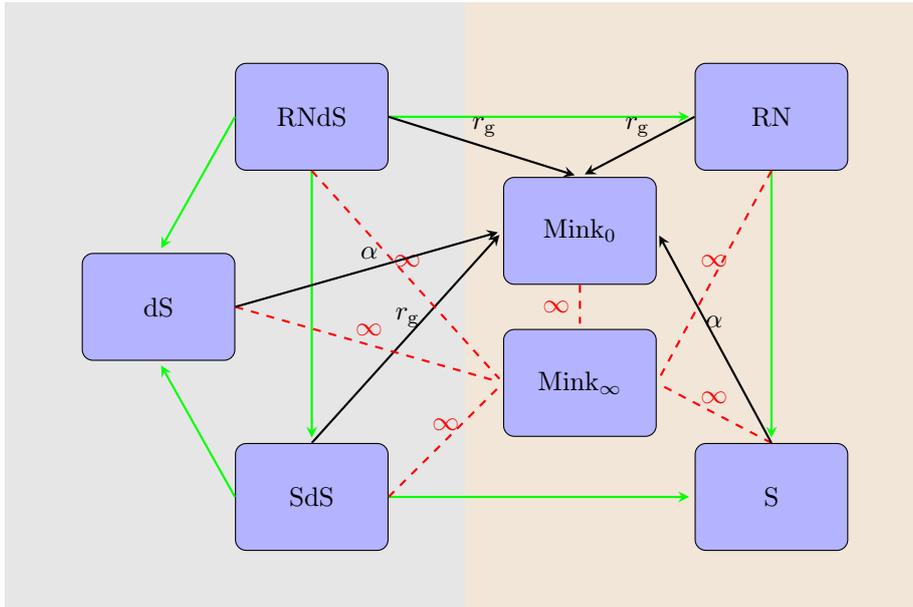
\begin{figure}[t!]
    \centering
   \begin{adjustbox}{width=0.8\textwidth}
\centering
 \begin{tikzpicture}
        \node[] (8) at (3,4.5) {\textbf{Asymptotically de Sitter}};
        \node[] (9) at (9,4.5) {\textbf{Asymptotically flat}};
        \fill[gray!20!white] (0,-4) rectangle (6,4);
        \fill[brown!20!white] (6,-4) rectangle (12,4);
       
		\node [block] (0) at (4, 2.5) {RNdS};
		\node [block] (1) at (2, 0) {dS};
		\node [block] (2) at (4, -2.5) {SdS};
		\node [block] (4) at (10, 2.5) {RN};
		\node [block] (6) at (10, -2.5) {S};
		\node [block] (7) at (7.5, 1) {Mink$_0$};
		\node [block] (10) at (7.5, -1) {Mink$_\infty$};
		
            \draw[->,green,thick,shorten >=2pt,>=stealth] (0.east) -- (4.west) node[midway,above] {}; 
            \draw[->,green,thick,shorten >=2pt,>=stealth] (4.south) -- (6.north) node[midway,right] {};
            \draw[->,green,thick,shorten >=2pt,>=stealth] (2.east) --(6.west) node[midway,above] {};
            \draw[->,green,thick,shorten >=2pt,>=stealth] (2.west)--(1.south) node[midway,left] {};
            \draw[->,green,thick,shorten >=2pt,>=stealth] (0.south)--(2.north) node[midway,left] {};
            \draw[->,green,thick,shorten >=2pt,>=stealth] (0.west)--(1.north)node[midway,right] {};
            \draw[->,black,thick,shorten >=2pt,>=stealth] (0.east)--(7.north) node[midway,above] {$r_\geo$};
            \draw[->,black,thick,shorten >=2pt,>=stealth] (4.west)--(7.north)node[midway,above] {$r_\geo$};
            \draw[->,black,thick,shorten >=2pt,>=stealth] (2.north)--(7.west) node[midway,above] {$r_\geo$};
            \draw[dashed,red,thick,shorten >=2pt,>=stealth] (1.east)--(10.west) node[midway,above] {$\infty$};
            \draw[dashed,red,thick,shorten >=2pt,>=stealth] (6.north)--(10.east) node[midway,above] {$\infty$};
             \draw[dashed,red,thick,shorten >=2pt,>=stealth] (7.south)--(10.north) node[midway,left] {$\infty$};
              \draw[dashed,red,thick,shorten >=2pt,>=stealth] (4.south)--(10.east) node[midway,above] {$\infty$};
               \draw[dashed,red,thick,shorten >=2pt,>=stealth] (2.east)--(10.west) node[midway,above] {$\infty$};
                \draw[dashed,red,thick,shorten >=2pt,>=stealth] (0.south)--(10.west) node[midway,above] {$\infty$};
            \draw[->,black,thick,shorten >=2pt,>=stealth] (1.east)--(7.west) node[midway,above] {$\alpha$};
            \draw[->,black,thick,shorten >=2pt,>=stealth] (6.north)--(7.east) node[midway,above] {$\alpha$};

\end{tikzpicture}
   
\end{adjustbox}
    \caption{Schematic summary of the moduli space distances between space-times. Dashed lines represent infinite distances, while solid lines stand for finite minimal distances. Mink$_\infty$ lies at an infinite distance of all other space-times.  We show in black the geodesic paths leading to Mink$_0$ and in green the paths discussed in Sec.~\ref{sec:Distances_between_space-times_withnonzeromass}.  }
    \label{fig:Conclusion_scheme}
\end{figure}

\section{Generalizing the Black Hole Entropy Distance Conjecture}
\label{sec:BHEDC}
In Sec.~\ref{sec:Distance} we analyzed the distances between space-times. We have determined that there always exist at least a minimal geodesic that connects the different space-times between them and to the zero mass limit Minkowski space-time Mink$_0$. Further, we have also identified a point in moduli space Mink$_\infty$ that represents the Minkowski limit when the mass of the black hole grows infinitely. The next step towards the generalized Black Hole Entropy Distance Conjecture lies in determining whether the distance in moduli space can always be expressed in terms of the total entropy of the space-time, as it was originally done in~\cite{Bonnefoy:2019nzv}.

In order to relate the distance in field space to the entropy of the space-time, we first need to define what we mean with "the entropy of the space-time", since only the entropy of a horizon is a priory unambiguously defined. Hence, in asymptotically flat space-times, the entropy of the space-time is defined just as a quarter of the area of the exterior black hole horizon. This fact reflects that an observer outside of the black hole only has access to the micro-state counting of the degrees of freedom of the black hole, regardless of the presence of a second Cauchy horizon inside the black hole\footnote{As is the case for a Reissner-Nordstr\o m black hole or the Reissner-Nordstr\o m-de Sitter space-time.}, undetectable for the exterior observer. 

Nevertheless, in asymptotically de Sitter space-times, we are forced to consider observers between two horizons. Further, generally the two horizons radiate at different temperatures and hence form a non-equilibrium system, for which a well defined concept of temperature does not exist. 

Hence, in the following we briefly review some known examples where the temperature and the entropy of the space-time can be defined unambiguously. We then extend this concept to non-equilibrium systems to suggest an upper bound for the entropy of the space-time. 

\subsection{Entropy of multi-horizon space-times}
\label{sec:Entropy}

The entropy of asymptotically flat space-times has been extensively discussed, it is given by a quarter of the area of the exterior black hole horizon~\cite{Bekenstein:1972tm,Hawking:1974sw}. The entropy of asymptotically de Sitter space-times has also been tackled, and exact results have been obtained for space-times in thermal equilibrium~\cite{Mann:1995vb,Hawking:1995ap}. Nevertheless, how to estimate the total entropy of space-times in absence of thermal equilibrium is still under debate. Usually, the total entropy of the space-time, that we  denote by $\mathcal{S}$, is just taken to be the sum of the entropies of the horizons surrounding the observer~\cite{Dinsmore:2019elr,Dolan:2013ft}. Here we argue that this estimate is an upper bound to the total entropy of the space-time. 

For the sake of completeness, we start by briefly reviewing the results of~\cite{Mann:1995vb,Hawking:1995ap} regarding the total entropy of space-times in thermal equilibrium. 

In order to define the thermodynamic quantities of a bifurcated horizon we work in the Euclidean path integral approach. Then, following the prescription of~\cite{Mann:1995vb},  we firstly analytically continue the metric of the space-time to the Euclidean sector by performing the transformation $t\to i\tau$. Next, we identify the  periodicity of the Euclidean time such that all conical singularities are removed from the metric. Notice that in asymptotically de Sitter space-times, we are forced to consider observers between the (exterior) black hole and the cosmological horizons. Hence, in general we have to deal with conical singularities at $r=r_\cc$ and $r=r_+$, which cannot be removed simultaneously~\cite{Choudhury:2004ph}.  Nevertheless, there exist some special cases within the physical phase space of the RNdS and the SdS space-times such that some of the horizons  radiate at the same temperature, and consequently, such that the conical singularities of the Euclidean section of the metric can be removed by choosing a certain periodicity for the Euclidean time. We refer to these  space-times as being in thermal equilibrium. 

The space-time in the Euclidean section describes a micro-canonical ensemble, and hence a closed thermodynamic system at fixed energy~\cite{Mann:1995vb}. The partition function of such an ensemble is given by $  Z = e^{-2I}$,  where $I$ is the Euclidean action~\cite{Mann:1995vb}
\begin{equation}\label{eq:ActionGeneral}
    I = -\frac{1}{16\pi} \int_M \dd^4 x \sqrt{g} (R -2\Lambda -F^2) +\frac{1}{8\pi} \int_{\Sigma} \dd^3 x \sqrt{\gamma} K \,,
\end{equation} and $K$ is the trace of the extrinsic curvature of the hypersurface $\Sigma$ with induced metric $h$.

The partition function $Z$ shall be interpreted as the density of states, so the entropy is just
\begin{equation}
    S=\log Z = -2I\,.
\end{equation}

There is a unique Schwarzschild-de Sitter space-time in thermal equilibrium, the Nariai limit\footnote{The discriminant locus~\eqref{delta} with zero charge reads
\begin{equation*}
    D = \frac{4}{\ell^4} (\ell^2-27 M^2)\,,
\end{equation*}
so the degenerate case occurs when the mass and de Sitter radius are not independent, but satisfy $\ell = 3\sqrt{3} M$.}.
Notice that for the conical singularities at $r=r_{+,\cc}$ to be simultaneously removed, the surface gravity of both horizons need to coincide, thus yielding a vanishing temperature of the horizons in the Nariai limit $T_\cc=T_+ = 0$. Hence, the entropy of the ensemble is given by~\cite{Hawking:1973qla} 
\begin{equation}\label{eq:StotalNariai}
   \mathcal{S}  = 2\pi r_\geo^2 = 2\frac{A_\geo}{4}\,,
\end{equation}
where $r_\geo$ is the position of the geodesic observer~\eqref{eq:rgeoSdS}.

The RNdS space-time presents four configurations in thermal equilibrium~\cite{Romans:1991nq}.  The explicit computation of the entropy of these space-times can be found in~\cite{Mann:1995vb}. Firstly,  the \textit{lukewarm} solution ($M=|Q|$ line in Fig.~\ref{fig:sharkfinxy}) is such that the exterior black hole and the cosmological horizons are in thermal equilibrium at finite temperature ($T_+=T_\cc$). Then, the conical singularities at $r=r_{+,\cc}$ can be removed by identifying the Euclidean time periodicity with $\tau \sim \tau + \frac{2\pi}{\kappa_{+,\cc}}$. This identification yields
\begin{equation}\label{eq:Slukewarm}
    \mathcal{S} =\pi \ell(\ell-2M) = \frac{1}{4}\left(A_+ + A_\cc\right)\,.
\end{equation} for  the total entropy for the lukewarm space-time. Next, the \textit{cold or extremal} space-time (upper branch in Fig.~\ref{fig:sharkfinxy}) is characterized by a degenerated black hole horizon at zero temperature. The degenerate horizon at $r=r_-=r_+$ is at an infinite proper distance along the space-like directions of any point $r$ in the space-time. Hence, the cold Euclidean metric only presents one conical singularity close to $r=r_\cc$ that can be avoided by identifying $\tau \sim \tau + \frac{2\pi}{\kappa_\cc}$, which yields
\begin{equation}
    \mathcal{S} = \pi r_\cc^2 =\frac{A_\cc}{4}\,. 
\end{equation} Similarly, the \textit{ultra-cold}  space-time~\eqref{eq:ultra-cold} presents a triple degenerated horizon at null temperature. Hence, expressing the  Lorentzian ultra-cold metric as $\text{Mink}^{(1,1)}\times S^2$ in Rindler coordinates yields
 \begin{equation}
   \mathcal{S} = \frac{2\pi}{\Lambda}\,,
\end{equation}
while using the usual coordinates leads to the $\text{Mink}^{(1,1)}\times S^2$ topology for the Lorentzian section~\cite{Mann:1995vb} and zero entropy.
 
Nevertheless, in absence of thermal equilibrium, there is at least one conical singularity that cannot be removed from the Euclidean section of the metric. Hence, the Euclidean metric cannot be made regular and the concept of temperature is ill-defined. This issue can be avoided by isolating the black hole and and the cosmological horizons into two distinct ensembles. This procedure was originally described by Hawking~\cite{Gibbons:1976ue} and further explored in~\cite{Saida:2011vu}.

In order to do so, we place an imaginary perfectly reflecting wall at $r_\cc \geq r_B \geq r_+$. The observer is located at $r_B$, so it can measure the state variables of both ensembles. After a finite period of time, the isolated regions $ r\in[r_+,r_B]$ and $r\in [r_B,r_\cc]$, that we denote by $\mathcal{E}_+$ and $\mathcal{E}_\cc$, reach the thermal equilibrium with the corresponding horizons. Hence, the periodicity of the Euclidean time is set in each ensemble as to avoid the conical singularity of the enclosed horizon, i.e.,  $\tau\sim \tau+\beta_*^{+} = 2\pi/\hat{\kappa}_+$ for $\mathcal{E}_+$ and $\tau\sim \tau+\beta_*^{\cc} = 2\pi/\hat{\kappa}_\cc$ for $\mathcal{E}_\cc$. Here we denoted by $\hat{\kappa}_h = \frac{1}{2} |V'(r_h)|$ the surface gravity normalized at infinity.  Further, the temperatures of the canonical ensembles $\mathcal{E}_{+,\cc}$ can be computed to be
\begin{equation}\label{eq:BetaEnsembles}
    \beta_+ = T_+^{-1} =\int_0^{\beta_*^{+}} \dd \tau \sqrt{g_{E\tau \tau}} = \frac{2\pi}{\hat{\kappa_+}} \sqrt{V(r_B)}\,, \quad  \beta_\cc = T_\cc^{-1} =\int_0^{\beta_*^{c}} \dd \tau \sqrt{g_{E\tau \tau}} =\frac{2\pi}{\hat{\kappa}_\cc} \sqrt{V(r_B)}\,,
\end{equation}
with $V(r_B)$ the SdS or RNdS metric function evaluated at the position of the heat wall.

The Euclidean action and the Helmholtz free energy of the ensembles $\mathcal{E}_+$ and $\mathcal{E}_\cc$ can be computed analogously to~\cite{Saida:2011vu}. The thermodynamic state variables follow by using the usual definitions of entropy and energy~\cite{York:1986it}. As discussed in~\cite{Saida:2011vu}, the total entropy of the space-time can be evaluated numerically, thus yielding
 \begin{equation}
     \mathcal{S} \sim \pi(r_+^2 + r_\cc^2) + f(r_+,r_\cc,r_B)\,,
 \end{equation} where $f(r_+,r_\cc,r_B)$ measures the discrepancy with respect to the usual definition of the total entropy of the space-time when there is thermal equilibrium\footnote{The detailed computation yielding this result will be presented elsewhere.}. This function $f(r_+,r_\cc,r_B)\leq 0$ is negative for arbitrary values of $r_+$, $r_\cc$ and $r_B$ and becomes strictly zero when we consider the thermal wall over one of the horizons $f(r_+,r_\cc,r_B=r_i)=0$ with $i=\{+,\cc\}$\footnote{This result is compatible with the construction presented in~\cite{Dinsmore:2019elr}, where a thermodynamic ensemble for the SdS space-time is built by considering the cosmological horizon as a thermal bath. The total entropy of the ensemble is then defined as the sum of the entropies of the black hole and cosmological horizon.}.  Therefore, this total entropy estimate is bounded by
 \begin{equation}\label{eq:Entropybound}
     \mathcal{S}\leq \pi(r_+^2 + r_\cc^2)\,.
 \end{equation}

\subsection{Generalized entropy-distance relation}

After having discussed the entropy of multihorizon space-times, we can now proceed and relate the entropy to the distance between them.
In the following we will denote the entropy of a space-time by $\mathcal S$.
In this section we wish to generalize the entropy-distance relation
\begin{equation}\label{eq:entropy-dist-relation}
    \Delta \sim  \log \mathcal S
\end{equation}
found for some restricted cases in~\cite{Bonnefoy:2019nzv} to arbitrary space-times that belong to the RNdS family.
Since the entropy of a space-time is not well-defined for systems out of equilibrium, we will test \eqref{eq:entropy-dist-relation} case by case.
Our results are summarized in table~\ref{tab:SummaryDistanceEntropy}.

\begin{table}
\centering
\begin{tabular}{ c c c c }
\hline
\multicolumn{1}{|c}{\textbf{Path}} & \textbf{$\alpha$} & \textbf{Entropy} & \multicolumn{1}{c|}{$\Delta\sim\log \mathcal S$}\\
\hline
Weyl rescaling & $\sqrt{\mathcal S}$ & $\mathcal S$ & $\checkmark$\\
\hline
$\{$ RNdS, SdS, RN $\}$ $\to$ Mink$_0$ & $r_\geo$ & ? & $(\times)$ \\
Nariai $\to$ Mink$_0$ & $r_\geo$ & $2 \pi r_\geo^2$ & $\checkmark$ \\
S $\to$ Mink$_0$ & & $4\pi M^2$ & $\times$ \\
dS $\to$ Mink$_0$ & & $\pi\ell^2$ & $\times$ \\
\hline
RN $\to$S & $r_+$ & $\pi r_+^2$ & $\checkmark$ \\
$\{$SdS, RNdS $\}$ $\to$ S & $r_+$ & ? & $(\checkmark)$\\
\hline
RNdS $\to$ RN & $\sqrt{M+\sqrt{M^2-Q^2}}$ & ? & $(\checkmark)$ \\
\hline
$\{$ RNdS, SdS$\}$ $\to$ dS & $r_\cc$ & ? & $(\checkmark)$\\
\hline
RNdS $\to$ SdS & $M$, $\ell$ & ? & ? \\
\hline
\end{tabular}
\caption{Summary table of the entropy-distance relation for different geodesic paths.
The first column denotes the path under consideration.
In the second column, the convenient choice for $\alpha$ is presented, if applicable.
The third column summarizes the relevant entropy in each case.
The last column shows whether the entropy-distance relation holds true, as indicated by $\checkmark$.
The symbol $(\checkmark)$ means that the relation holds true in the limit while $(\times)$ indicates inconclusive results.}
\label{tab:SummaryDistanceEntropy}
\end{table}

\subsubsection{Entropy-distance relation in the Mink${}_\infty$--limit}

The limit $\text{Mink}_\infty$ is reached for proper time $\lambda\to\infty$ and hence is at infinite distance for every parametrization $\alpha$.
For simplicity we first restrict ourselves to those parametrizations such that the metric flow corresponds to Weyl rescalings.
In this case, all mass parameters depend on $\lambda$ only exponentially as in Eq.~\eqref{eq:mqloflambda} for $c_M=c_Q=c_\ell=0$.
First, we focus on space-times where an observer has access to only one horizon, i.e. S, dS and RN.
Let $r_h$ denote the radial position of a horizon, then $r_h \propto e^{\lambda/4}$ due to dimensional reasons.
Hence, the choice $\alpha=r_h$ corresponds to Weyl rescalings.
In these cases the entropy of the space-time is unambigously given by the Bekenstein-Hawking formula $\mathcal S = 4 \pi r_h^2$.
It immediately follows the relation $\alpha = \sqrt{\mathcal S/4\pi}$ and hence Eq.~\eqref{eq:entropy-dist-relation}. 
Turning to the case where the observer outside the black hole has access to two horizons, i.e. SdS and RNdS, we can draw the same conclusion.
The entropy of the space-time must flow with $\lambda$ as $\mathcal S \propto e^{\lambda/2}$ again by dimensional reasons.
Hence, the parametrization $\alpha \propto \mathcal S^{1/2}$ represents Weyl rescalings also in these cases.
This establishes the entropy-distance relation~\eqref{eq:entropy-dist-relation} even if the precise form of $\mathcal S$ is not known for these multihorizon space-times.
We conclude that the entropy-distance relation~\eqref{eq:entropy-dist-relation} holds true for every space-time configuration belonging to the RNdS family in the case of Weyl rescalings.
Let us emphasize that Eq.~\eqref{eq:entropy-dist-relation} holds true in general and not only in the limit Mink${}_\infty$.

\subsubsection{Entropy-distance relation in the Mink${}_0$--limit}

We proceed by testing the entropy-distance relation in the limit of Mink${}_0$.
For the cases where an observer has access to only one horizon that we denote by $r_h$, i.e. S, dS and RN, the parametrization $\alpha=r_h$ leads to an infinite distance.
As in the case of Mink${}_\infty$, the Bekenstein-Hawking formula immediately yields $\alpha \propto \mathcal S^{1/2}$.
Therefore, Eq.~\eqref{eq:entropy-dist-relation} also holds true in the zero entropy limit.
However, other parametrizations are possible, which do not correspond to Weyl rescalings and yield finite distance.
The entropy-distance relation cannot be established for these types of flows.

Moving on to cases where the observer has access to two horizons, i.e. SdS and RNdS, we have established $\alpha=r_\geo$ as appropriate parametrization.
To relate $r_\geo$ to the entropy, we first discuss those multihorizon space-times for which a notion of entropy exists, i.e. space-times in thermal equilibrium.
As discussed in Sec.~\ref{sec:Entropy}, in the case of Nariai black hole, as well as extremal RN black hole (for an observer between the interior and exterior black hole horizons), the total entropy can be written as $\mathcal S = 2 A_\geo/4 = 2\pi r_\geo^2$.
This indeed implies Eq.~\eqref{eq:entropy-dist-relation} to be exact for these cases.
For the lukewarm, cold and ultra-cold space-time however, there is no geodesic to Mink${}_0$ that maintains thermal equilibrium.
Hence, the entropy-distance relation cannot be tested in these cases.

Finally, we study the entropy-distance relation for space-times out of thermal equilibrium, where the observer has access to two horizons, i.e. SdS, RN, RNdS.
The geodesic observer yields the convenient parametrization here, $\alpha=r_\geo$, so that the distance of these space-times to Mink${}_0$ is given by
\begin{equation}\label{eq:DistanceRg}
    \Delta = 4c \log\left(\frac{r_{\rm g,f}}{r_{\rm g,i}}\right)\,.
\end{equation}
Since the limit is arrived at after a finite amount of proper time, the distance is finite.
The entropy-distance relation~\eqref{eq:entropy-dist-relation} would hold in these cases, if the entropy of these space-time was related to the geodesic observer as $\mathcal S=2\pi r_\geo^2$.
As an example, consider the path SdS $\to$ Mink${}_0$. The total entropy of SdS needs to satisfy Eq.~\eqref{eq:Entropybound}. Now we test whether the estimate $\mathcal S=2\pi r_\geo^2$ for the total entropy of SdS is a good ansatz.  Expressing this ansatz in terms of the individual entropies of the black hole horizon $S_+$ and the cosmological horizon $S_\cc$ yields
\begin{equation}
    \mathcal S = 2\left(\frac{1}{4} S_+ S_\cc (S_+ + S_\cc + 2\sqrt{S_+ S_\cc})\right)^{1/3}
\end{equation}
This ansatz respects the inequality $\mathcal S \leq S_{+} + S_\cc$ as we expect for the total entropy of a multi-horizon space-time.
However, in the limit without black hole, i.e. pure dS, our ansatz does not reduce to the entropy of the de Sitter horizon, but instead $\mathcal S\rightarrow 0$ as $\mathcal S_+\rightarrow 0$.
We conclude that $2\pi r_\geo^2$ does not correspond to the entropy of multi-horizon space-times out of thermal equilibrium.
Although this indicates that the entropy-distance relation cannot be established in these cases, we would need to test the distance~\eqref{eq:DistanceRg} against the analytical expression of the total entropy of these space-times to have a fully conclusive result.

\subsubsection{The BHEDC for space-times with non-zero mass parameters}

Here we briefly identify whether the geodesic distance between the space-times with non-zero mass parameters described in Sec.~\ref{sec:Distances_between_space-times_withnonzeromass} can be related to the total entropy of the space-time. 

We identified a geodesic connecting the RN and to S parametrized by $\alpha= r_{+}$.
Therefore, it follows that the distance between these space-times can be expressed as a function of the total entropy of the space-time
\begin{equation}
    \Delta = 2c \log\left(\frac{\mathcal S_{\rm f}}{\mathcal S_{\rm i}}\right)\,.
\end{equation}
Here, the final entropy is the one of Schwarzschild space-time, i.e. $\mathcal S_{\rm f} = 4\pi M_{\rm f}^2$, and the inital entropy is the one of RN for an observer outside the black hole, i.e.  $\mathcal S_{\rm i} = \pi(M_{\rm i}+\sqrt{M_{\rm i}^2-Q_{\rm i}^2})^2$. 

For the paths connecting SdS and RNdS to S there is no well-defined notion of entropy.
However, in the Schwarzschild-limit of these paths, the parametrization becomes proportional to the mass of the black hole, $\alpha\sim M$.
This implies that $\alpha\propto\mathcal S^{1/2}$ where $\mathcal S = 4\pi M^2$ asymptotically.
Hence, the entropy-distance relation~\eqref{eq:entropy-dist-relation} holds true in the limit.

\section{Application to black hole evaporation}
\label{sec:BHEvaporation}
Finally, we apply our previous findings to the evaporation of black holes.
We want to test whether a toy model of black hole evaporation passes the known Swampland criteria~\cite{Brennan:2017rbf,Palti:2019pca}.

As such, we consider a black hole with charge because global symmetries are not allowed~\cite{Harlow_2019,harlow2019symmetries}.
We thus study the Reissner-Nordstr\o m-de Sitter black hole  here.
Classically, it is stable~\cite{Zhang:2019nye},
but it can decay when quantum effects are taken into account.
Specifically, the black hole can loose charge and mass via the Schwinger effect~\cite{Schwinger:1951nm} while both the black hole and cosmological horizon give rise to Hawking radiation~\cite{Hawking:1974sw,Gibbons:1976ue,Gibbons:1977mu}.

The discharge of a black hole through the combined effect of Hawking and Schwinger can be studied in two regimes.
These are controlled by the Schwinger transition rate,
\begin{equation}
    \Gamma \sim e^{-\frac{m^2}{q E}}\,,
\end{equation}
where $m$ and $q$ are the  mass and charge of the particle produced by the Schwinger effect and $E$ is the electric field due to the charged black hole.
In the regime of adiabatic discharge where $m^2\ll qE$ the Schwinger effect is exponentially enhanced and the black hole can loose all its charge very rapidly.
For initial black holes close to the charged Nariai branch the adiabatic discharge evolves the space-time towards a superextremal neutral Nariai solution, which is outside the phase space $\mathcal D$.
Hence, this regime evolves the space-time from a physical to an unphysical solution in the sense that the Cosmic Censorship Conjecture gets violated dynamically.
The opposite regime is the quasi-static discharge where $m^2\gg qE$. Since the Schwinger production rate is exponentially suppressed, the discharge happens very slowly.
As a result, any space-time that is initially within the phase space $\mathcal D$ remains within the phase space during the entire evaporation process.
In this sense, the quasi-static discharge connects physical solutions of Einstein's equation with each other.
This observation motivated the authors of Ref.~\cite{Montero:2019ekk} to formulate a de Sitter version of the Weak Gravity Conjecture (WGC)~\cite{ArkaniHamed:2006dz}: every particle in the spectrum must satisfy $m^2> qg M_P H$ in order to avoid super-extremality.
In the following, we take the WGC of~\cite{Montero:2019ekk} as basis such that the black hole discharges quasi-statically.

Finally, we consider the cosmological constant to be nonzero and positive to make contact with our Universe.
To respect the de Sitter conjectures~\cite{Dvali:2014gua,Dvali:2017eba,Obied:2018sgi,Dvali:2018fqu,Garg:2018reu,Ooguri:2018wrx,Dvali:2018jhn}, we allow the cosmological constant to vary in time.
On time scales on which the mass and charge of the black hole vary significantly\footnote{We take the time scale to be of the order of the geometric time scale, i.e.  $\tau\sim M$  in Hubble units.
As a first order estimate, a quasi-static process would take $\delta \tau \sim 510$ million years for the mass of the black hole to vary $\delta M/M \sim 0.1$ for sufficiently massive black holes.}, the drift of the cosmological constant has to be taken into account~\cite{Peebles:1999ie}.

It remains to test black hole evaporation against the Distance Conjecture~\cite{Ooguri:2006in} and in particular the Black Hole Entropy Distance Conjecture~\cite{Bonnefoy:2019nzv}, which we generalized in the previous section to apply it to evaporation.
This is the subject of the present section.

We work out the technical details in appendix~\ref{sec:appC}.
We closely follow the analysis of~\cite{Montero:2019ekk}, but extend to a time-varying cosmological constant.
We first identify the endpoint of the evaporation process and present the corresponding trajectories in the phase space $\mathcal D$.
We then emulate these trajectories by moduli space geodesics.

%%%%%%%%%%%%%%%%%%%%%%%%%%%%%%%%%%%%%%%%%%%%%%%%%%%%%%%%
\subsection{Phase space trajectories of evaporation}
%%%%%%%%%%%%%%%%%%%%%%%%%%%%%%%%%%%%%%%%%%%%%%%%%%%%%%%%

We discuss the evaporation trajectories first for a general RNdS black hole and then specify to SdS and Nariai.
We leave all technical details to appendix~\ref{sec:appC} as we closely follow the perturbative procedure of~\cite{Montero:2019ekk}.
The accumulated effect of Hawking and Schwinger radiation back-reacts on the geometry in a quasi-static way such as to preserve staticity and spherical symmetry.
In other words, we restrict ourselves to a self-similar evaporation.

We extend the analysis of~\cite{Montero:2019ekk} by allowing the cosmological constant to vary in time.
We use the first law of black hole mechanics with multiple horizons as presented in Ref.~\cite{Dolan:2013ft}\footnote{We
derive the dynamics of the cosmological constant from the first law of thermodynamics in asymptotic de Sitter space-times. The cosmological constant needs to be considered a state variable, which enters the first law as a pressure term conjugated to the volume of the space-time~\cite{Sekiwa:2006qj,Dolan:2013ft}. Although classically the laws of thermodynamics are only applicable to stationary black holes~\cite{Wald:2002mon}, using the first law of thermodynamics as a dynamical equation is justified for the following reasons. First, we consider a quasi-static regime of evaporation. Second, the formalism used to derive the first law of thermodynamics in asymptotically de Sitter space-times in~\cite{Dolan:2013ft} is equivalent to treating one of the horizons as a boundary~\cite{Gomberoff:2003ea}. Hence, the first law derived in~\cite{Dolan:2013ft} fits into the formalism of Isolated Horizons~\cite{Ashtekar:1999yj,Ashtekar:2004cn}, which in turn justifies that the thermodynamic equations are enough to consider a consistent Hamiltonian evolution~\cite{Ashtekar:2000sz}.}.

In this section we discuss physical evaporation processes.
We therefore restore the constants that we have set to unity in the beginning.
We normalize the length scales to the value of the cosmological horizon today, $\ell_0 = 1.6\times 10^{26}\,\text{m}$.
This amounts to define the dimensionless radial coordinate $\tilde r = r / \ell_0$ as well as the dimensionless mass parameters 
\begin{equation}
\tilde M = \frac{M}{\ell_0} = \frac{Gm}{c^2\ell_0}\,, \quad
\tilde Q = \frac{Q}{\ell_0} = q\sqrt{\frac{g_0^2G}{4\pi\ell_0^2 }}\,, \quad
\tilde \ell = \frac{\ell}{\ell_0}\,.
\end{equation}
In the rest of this section, we will suppress the tilde for the sake of visibility.

%%%%%%%%%%%%%%%%%%%%%%%%%%%%%%%%%%%%%%%%%%%%%%%%%%%%%%%%
\subsubsection{Evaporation of RNdS black hole}
\label{sec:RNdSEvaporation}
%%%%%%%%%%%%%%%%%%%%%%%%%%%%%%%%%%%%%%%%%%%%%%%%%%%%%%%%

As summarised in appendix~\ref{sec:summary-evaporation}, the system of differential equations can be decoupled.
It is convenient to introduce the variables $x= M/\ell$ and $y= Q/\ell$.
The system of differential equations can then be written as
\begin{align}\label{eq:dxdtdydt}
\begin{split}
\mathring{x} &= \frac{4\pi r_\geo^2}{\ell} \Bigg[\left(G\sqrt{ V(r_\geo)} +V(r_\geo)^2\frac{r_\geo^3-x \ell^3}{r_\cc^3-r_+^3}\right)\mathcal{T}_\geo - \left(\frac{Q}{r_\geo} + \frac{r_\geo^3-x \ell^3}{r_\cc^3-r_+^3}\left(\frac{Q}{r_+}-\frac{Q}{r_\cc}\right)\right) \mathcal{J}_\geo\Bigg]\\
\mathring{y} &= -\frac{4\pi r_\geo^2}{\ell} \Bigg[ \frac{y \ell^3}{r_\cc^3-r_+^3} V(r_\geo)^2\mathcal{T}_\geo  + \left(1-\frac{y \ell^3}{r_\cc^3-r_+^3} \left(\frac{Q}{r_+}-\frac{Q}{r_\cc}\right)\right) \mathcal{J}_\geo\Bigg]\,.
\end{split}
\end{align}
Here, the operator $\mathring{ } = \dd/\dd  t_\geo$ denotes derivative with respect to the proper time of the geodesic observer.
Further, the Hawking flux is given by
\begin{equation}
    \mathcal{T}_\geo = \frac{\sigma}{(4\pi)^3}\frac{1}{V(r_\geo)^2}(r_\text{c}^2|V'(r_\text{c})|^4-r_+^2|V'(r_+)|^4)
\end{equation}
as in Eq.~\eqref{Eq:Tflux}.
The expression for the Schwinger flux $\mathcal J_\geo$ is  too lengthy to display here, but can be found in Eq.~\eqref{Jflux}.

\begin{figure}
\centering
\includegraphics[width=0.8\textwidth]{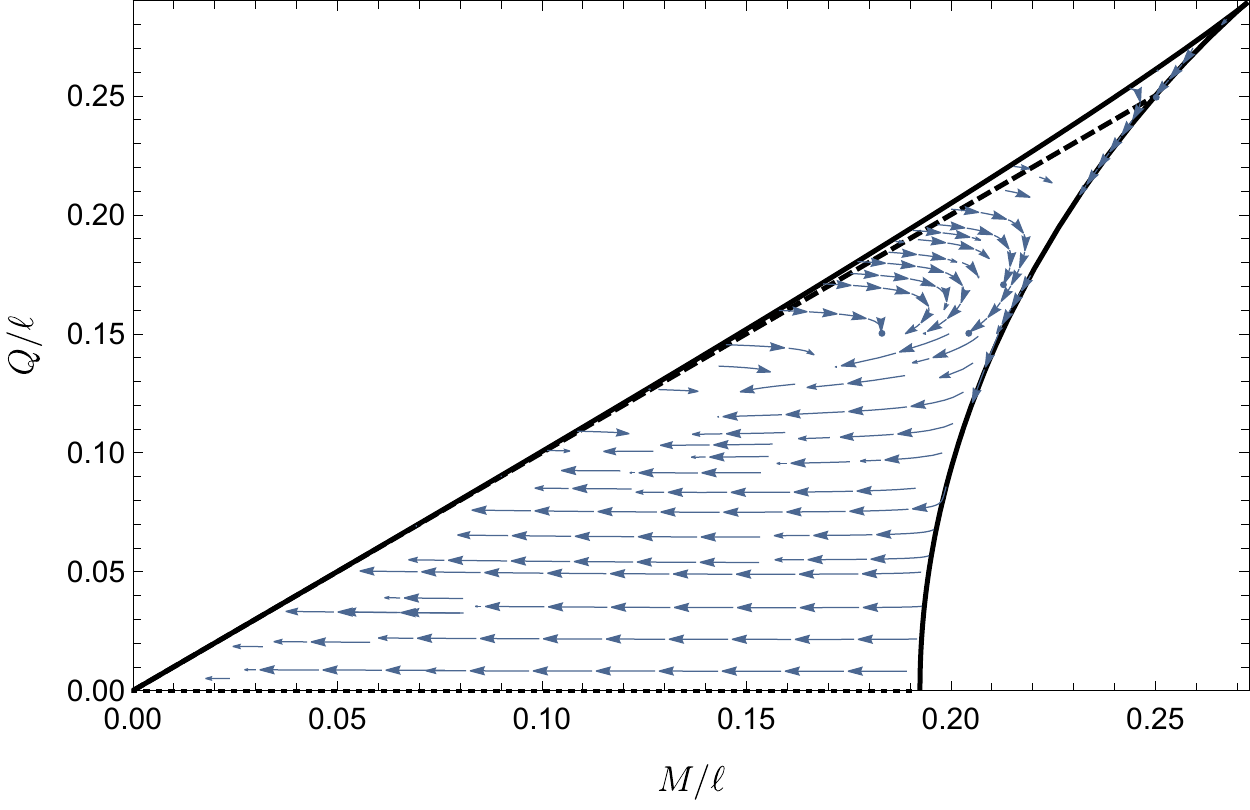}
\caption{Phase space evaporation of RNdS black holes in the $M/\ell$-$Q/\ell$-plane.
The black solid line represents the boundary of the physical phase space $\partial\mathcal D$.
The dotted black line indicates the lukewarm line.
The blue arrows represent the flow of the mass parameters due to the evaporation by Schwinger and Hawking effect.
While for large values of these ratios a period of anti-evaporation occurs, the space-time first looses mass and then evolves along the lukewarm line in the other regions of $\mathcal D$.
The endpoint of the evaporation process is de Sitter space-time filled with thermal radiation.}
\label{fig:dxdymodel2}
\end{figure}

This system of first order differential equations can be solved numerically for arbitrary initial space-time configurations within the phase space region $\mathcal{D}$. 
In Fig.~\ref{fig:dxdymodel2}, we depict the evaporation flow of the mass parameters of a RNdS space-time in the $M/\ell-Q/\ell$ phase space as dictated by Eq.~\eqref{eq:dxdtdydt}.
The boundary of the physical phase space $\partial \mathcal{D}$ and the lukewarm line are depicted in black solid and dotted, resp.
Two regimes are appreciable:
\begin{itemize}
\item 
For large values of $M/\ell$ and $Q/\ell$ a process of anti-evaporation takes place, i.e. the mass of the black hole increases while its charge decreases.
In this regime, the space-time slowly evolves in the direction of the Nariai branch.
The black hole is too large as to compensate the radiation coming from the cosmological horizon, that increases the mass of the black hole.
As the mass of the black hole increases, its temperature follows, until the black hole is hot enough as to revert the mass flux coming from the Hawking radiation.
However, the thermal equilibrium is never reached for initially non-Nariai space-times. 
This feature is expected from the third law of thermodynamics: the zero temperature limit cannot be reached in finite time.
Instead, the space-time slowly evolves towards an evaporation regime, where both the mass and charge decrease.

\item
For smaller charge and mass ratios, the mass parameters evolve toward the lukewarm line.
The mass loss dominates the process, while the de Sitter horizon increases.
In this regime the black hole temperature is higher than that of the cosmological horizon, so there is a net mass flux from the black hole to the cosmological horizon.
Once the lukewarm line is reached, the exterior black hole horizon and the cosmological horizon reach thermal equilibrium, so the net Hawking radiation cancels out.
Notice that the mass loss of the black hole is not only due to the emission of neutral pairs, but also due to Schwinger.
This is reflected in the fact that Eq.~\eqref{eq:dxdtdydt} presents a term $\mathring{x} \sim \mathcal{J}_\geo$.
Consequently, the black hole evaporation and discharge continues along the lukewarm branch until empty de Sitter space-time is reached.
This evaporation is driven solely by the Schwinger effect, as it can be seen in Eqs.~\eqref{eq:dxdtdydt}.
\end{itemize}

During the whole process of evaporation, the black hole losses its charge until the neutral limit is reached.
This is also true along the Nariai and extremal branches, since the Schwinger flux is well defined along the whole phase space $\mathcal{D}$, also along the Nariai branch.
Further, the Schwinger charge flux is negative along the whole physical phase space, so there is a net charge loss that only stops when the black hole depletes completely. 

The origin $(M/\ell,Q/\ell)=(0,0)$ not only represents empty de Sitter space, but also the family of Reissner-Nordstr\o m black holes in asymptotically flat space-time.
Our discussion in section~\ref{sec:NariaiEvaporation} will elucidate that the endpoint of evaporation is indeed empty de Sitter space-time.
Hence, we can summarize this section by stating:
\textit{initially non-Nariai space-times evolve to empty de Sitter space-time}.

%%%%%%%%%%%%%%%%%%%%%%%%%%%%%%%%%%%%%%%%%%%%%%%%%%%%%%%%
\subsubsection{Evaporation of Schwarzschild-de Sitter black holes}
\label{sec:NariaiEvaporation}
%%%%%%%%%%%%%%%%%%%%%%%%%%%%%%%%%%%%%%%%%%%%%%%%%%%%%%%%

Next, let us study the Schwarzschild-de Sitter (SdS) space-time by setting $Q = \dot{Q} = 0$.
We study the evaporation equations for the mass of the black hole and the cosmological radius according to the Gibbons-Hawking model presented in Appendix~\ref{sec:appC}.

\begin{figure}
\centering
\includegraphics[width=0.8\textwidth]{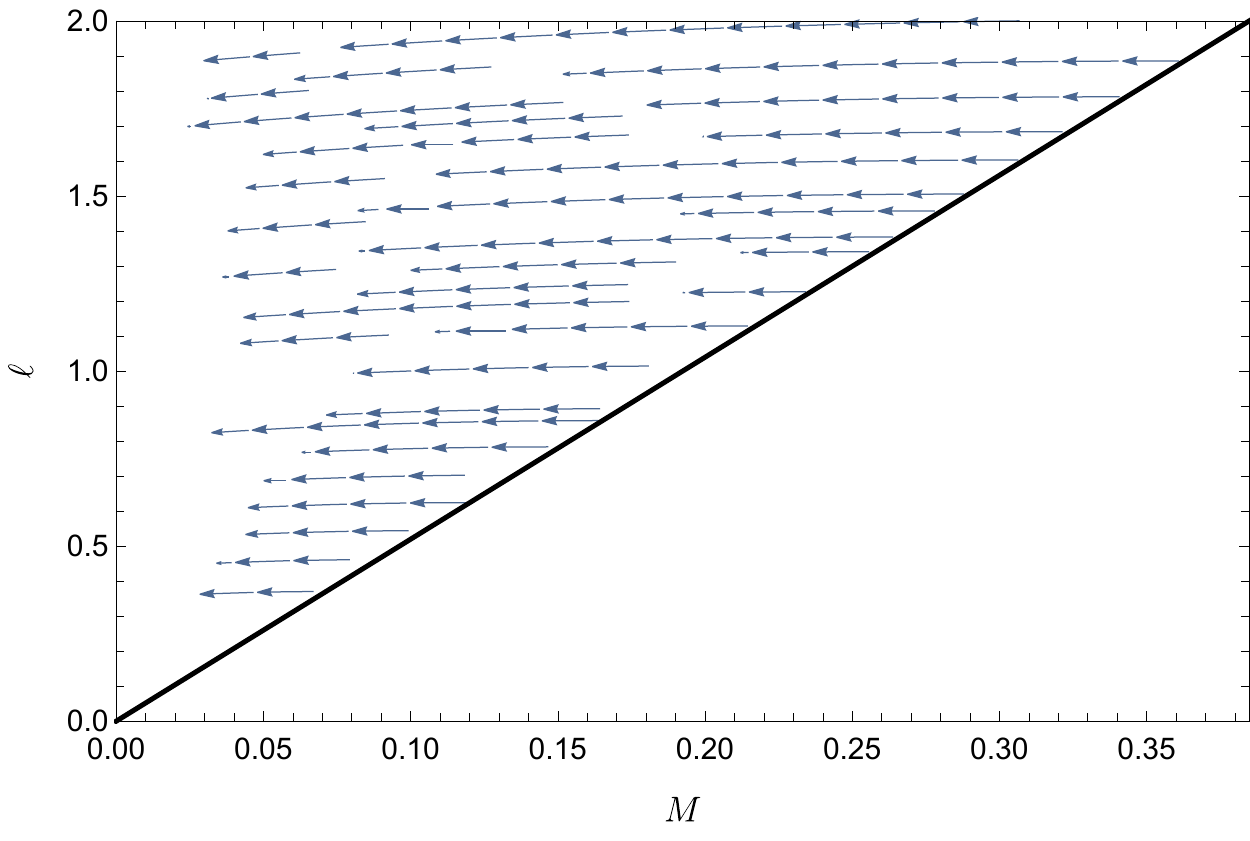}
\caption{Phase space evaporation of the Schwarzschild-de Sitter black holes in the $M-\ell$ plane. The Schwarzschild black hole fully evaporates to empty de Sitter space-time. The space-times initially in the Nariai limit do not evaporate.}
\label{fig:SdSevaporation2ndmodel}
\end{figure}

In the neutral limit, the evolution equations (\ref{eq:evaporationEqsModel2})
reduce to
\begin{align}\label{eq:SdSmodel2}
\mathring{M} = 4\pi r_\geo^2  \left(G\sqrt{ V(r_\geo)} + \frac{r_\geo^3 V(r_\geo)^2}{r_\cc^3-r_+^3}\right)\mathcal{T}_\geo\,,\quad\quad
\mathring{\ell} = 4\pi r_\geo^2\frac{\ell^3}{r_\cc^3-r_+^3} V(r_\geo)^2 \mathcal{T}_\geo\,.
\end{align}
The Schwinger effect is absent for neutral black holes.
The evolution of mass $M$ and de Sitter radius $\ell$ are depicted in Fig. (\ref{fig:SdSevaporation2ndmodel}).

Let us first discuss the Nariai branch, that is represented by the black solid line in Fig.~\ref{fig:SdSevaporation2ndmodel}.
Both the black hole and the cosmological horizons have the same temperature such that the system is in thermal equilibrium.
Hence, there is no net Hawking radiation between the horizons.
From Eqs.~\eqref{eq:SdSmodel2}, we see that the mass of the black hole and the cosmological constant remain constant in this limit.

However, the Nariai branch is unstable.
Space-times infinitely close to the Nariai branch do not evolve towards thermal equilibrium, but loose mass while $\ell$ remains almost constant.
As before, this behaviour is expected due to the third law of black hole mechanics, non-initially cold black holes cannot evolve towards cold ones in finite time. 

Instead, space-times that are initially away from the Nariai branch evolve towards empty de Sitter space-time, as can be seen from Fig.~\ref{fig:SdSevaporation2ndmodel}.
The de Sitter radius decreases slightly during the process, but it remains finite and positive. 
Therefore, SdS black holes evaporate completely. During this process the background becomes more accelerated due to the disappearance of the pull of the black hole.  

%%%%%%%%%%%%%%%%%%%%%%%%%%%%%%%%%%%%%%%%%%%%%%%%%%%%%%%%
\subsubsection{Evaporation of charged Nariai black hole}
%%%%%%%%%%%%%%%%%%%%%%%%%%%%%%%%%%%%%%%%%%%%%%%%%%%%%%%%

Finally, we discuss the evaporation of charged Nariai space-times.
As mentioned before, the integrated Schwinger flux \eqref{Jflux}  is not valid along the Nariai branch, for which $r_+\to r_\cc$. Nevertheless, the Schwinger flux is well defined along the whole phase space $\mathcal{D}$, including the Nariai branch. The integrated expression for the Schwinger flux along the Nariai branch can be obtained through instantonic methods, for instance see \cite{Montero:2019ekk,Frob:2014zka}. 
In this section we show that a Nariai space-time discharges along the Nariai branch until the neutral limit is reached. Firstly, notice that the charge flux is negative along the whole phase space $\mathcal{D}$. The charged Nariai branch is not stable although the black hole and the cosmological horizon are in thermal equilibrium. Initially Nariai space-times experience a mass and charge loss due to the Schwinger effect. The mass loss along the Nariai branch occurs due to the energy loss of the pairs produced when the charged carriers are repelled from the black hole horizon. 

Along the Nariai branch the evaporation equations~(\ref{eq:dxdtdydt}) simplify to
\begin{equation}\label{eq:dxdyNariai}
    \frac{\dd x}{\dd y} = \frac{y}{\tilde r_\geo} \frac{1+\frac{\tilde r_\geo^3-x}{\tilde r_\cc^3-\tilde r_+^3} \frac{\tilde r_\geo}{y}\left(\frac{y}{\tilde r_+} -\frac{y}{\tilde r_\cc}\right)}{1- \frac{y}{\tilde r_\cc^3-\tilde r_+^3}\left(\frac{y}{\tilde r_+} -\frac{y}{\tilde r_\cc}\right) }\,,
\end{equation}
because the Hawking flux vanishes, $\mathcal T_\geo=0$.
Here, the tildes indicate that the radial quantities are normalized by the de Sitter radius, i.e., $\tilde r_a = r_a/\ell$ for $a=\{+,\geo,\cc\}$.
From the definition of the geodesic observer $V'(r_\geo) =0$ it follows that $y^2 = \tilde r_\geo (x-\tilde r_\geo^3)$.
Plugging this into Eq.~(\ref{eq:dxdyNariai}) we find the following simple relation
\begin{equation}\label{eq:dxdynariaiSimplified}
    \frac{\dd x}{\dd y} = \frac{y}{\tilde r_\geo}\,.
\end{equation} 

Along the Nariai branch the black hole and cosmological horizon coincide with the position of the geodesic observer, $r_+=r_\cc = r_\geo$.
In this limit, the expression for the geodesic observer vastly simplifies,
\begin{equation}
\tilde r_\geo = \frac{1}{\sqrt{6}} \sqrt{1+\sqrt{1-12 y^2}}\,.
\end{equation}
Hence, we can integrate Eq.~(\ref{eq:dxdynariaiSimplified}) analytically to
\begin{align}
\begin{split}
x-x_0 & = \sqrt{6} \int_{y_0}^y \dd y' \frac{y'}{ \sqrt{1+\sqrt{1-12 y'^2}}} \\
& = \frac{1}{3\sqrt{6}} \left(\sqrt{1+36y_0^2+(1-12 y_0^2)^{3/2}}-\sqrt{1+36y^2+(1-12 y^2)^{3/2}}\right)\,,
\end{split}
\end{align} which coincides with the expression for the Nariai branch~\eqref{Eq:Nariai}.
This simple computation shows that initial Nariai space-times characterized by $(x_0,y_0)$ evolve along the Nariai branch, slowly discharging until the neutral Nariai limit is reached.

Finally, this result together with the evaporation paths in phase space for non-Nariai space-times shows that the physical evolution of RNdS space-time always maps physical space-times to physical space-times given our assumptions.

%%%%%%%%%%%%%%%%%%%%%%%%%%%%%%%%%%%%%%%%%%%%%%%%%%%%%%%%
\subsection{Evaporation paths and moduli space geodesics}
%%%%%%%%%%%%%%%%%%%%%%%%%%%%%%%%%%%%%%%%%%%%%%%%%%%%%%%%

We are finally in the position to relate the geometric distance between space-time configurations to the physical evaporation process that connects different space-times.
In this section, we will construct paths in the moduli space of RNdS space-times that solve the geodesic equation~\eqref{eq:geodesicEq} and emulate the physical evaporation process as presented in Fig.~\ref{fig:dxdymodel2}.
This allows to determine the geometric distance between any initial space-time configuration and its final state that is arrived at by a physical evaporation process.

%%%%%%%%%%%%%%%%%%%%%%%%%%%%%%%%%%%%%%%%%%%%%%%%%%%%%%%%
\subsubsection{Distance of evaporating Nariai black hole}
%%%%%%%%%%%%%%%%%%%%%%%%%%%%%%%%%%%%%%%%%%%%%%%%%%%%%%%%

As a warm-up, we first analyze the simple case of a charged Nariai black hole.
As discussed in Sec.~\ref{sec:NariaiEvaporation}, an initially Nariai space-time discharges following the Nariai branch until the SdS limit is reached.
Hence, we need to find a geodesic path whose projection on the phase space follows the Nariai branch. 
The problem reduces to finding an appropriate parametrization $\alpha$.
For Nariai space-times, the outer black hole and cosmological horizon coincide with the geodesic observer, $r_+ = r_\cc = r_\geo$.
Here we show that $\alpha= r_\geo$ parametrizes a geodesic along the Nariai branch.

Rearranging $V(r_\geo)=0$ we find that the mass and the charge are related to the location of the degenerate horizons or the geodesic observer as
\begin{equation} \label{Eq:BdryCondition}
M = r_\geo \left(1-\frac{2r_\geo^2}{\ell^2}\right), \quad Q^2 = r_\geo^2 \left(1-\frac{3r_\geo^2}{\ell^2}\right)\,.
\end{equation}
Hence, it is enough to show that the mass, charge and de Sitter radius evolution~\eqref{eq:mqloflambda} is compatible with Eq.~\eqref{Eq:BdryCondition} when $\alpha = r_\geo$.
Using the ansatz $r_\geo = r_{\geo,\rm i} e^{\lambda/4}$ and rearranging Eq.~\eqref{Eq:BdryCondition} yields
\begin{equation}
    M = M_{\rm i} e^{\lambda/4}\left(1-\frac{2 r_{\geo,{\rm i}}^2 c_\ell}{\ell_{\rm i}^2-2 r_{\geo,{\rm i}}^2}  \lambda\right)\,,\quad
    Q = Q_{\rm i} e^{\lambda/4} \sqrt{1 - \frac{3 r_{h,{\rm i}}^2 c_\ell}{\ell_{\rm i}^2-3 r_{\geo,{\rm i}}^2}\lambda}
\end{equation} where $M_{\rm i} = r_{\geo,{\rm i}}(1-2 r_{\geo,{\rm i}}^2/\ell_{\rm i}^2)$ and $Q_{\rm i}^2 = r_{\geo,{\rm i}} (1-3 r_{\geo,{\rm i}}^2/\ell_{\rm i}^2)$. Further, identifying the constants  $c_M=-\frac{2 r_{h,{\rm i}}^2 c_\ell}{\ell_{\rm i}^2-2 r_{h,{\rm i}}^2}$ and $c_Q = -\frac{3 r_{h,{\rm i}}^2 c_\ell}{\ell_{\rm i}^2-3 r_{h,{\rm i}}^2}$, we recover Eq.~\eqref{eq:mqloflambda}.

The initial mass, charge and de Sitter radius need to be related through Eq.~\eqref{Eq:Nariai}. The largest possible values for the ratios $x=M/\ell$ and $y=Q/\ell$ correspond to the ultra-cold point,
\begin{equation}\label{eq:Ultra-cold}
    \frac{M}{\ell} =\frac{1}{3}\sqrt{\frac{2}{3}}\,,\quad \frac{Q}{\ell} =\frac{1}{\sqrt{12}}\,.
\end{equation}
Hence, in order to define a geodesic which connects the ultra-cold point to the neutral Nariai limit we set $(x_{\rm i},y_{\rm i})$ to \eqref{eq:Ultra-cold}. This geodesic describes the complete discharge of the black hole along the Nariai branch. Hence, the proper time ranges over $\lambda\in [0,-1/c_Q]$ such that $Q(\lambda = -1/c_Q)=0$. This choice of constants fully specifies the geodesic connecting the ultra-cold and the neutral Nariai space-times.  

Summarizing, the parametrization $\alpha = r_\geo$ defines a geodesic that evolves Nariai solutions to Nariai solutions.
The neutral Nariai black hole is at finite distance from the ultra-cold Nariai black hole (as well as any other charged Nariai black hole) when following the physical evaporation trajectory.
During this process of evaporation the horizon radius $r_\geo$ increases monotonically.
With the parametrization $\alpha = r_\geo$, Eq.~(\ref{eq:distance-final}) implies for the geometric distance
\begin{equation}
    \Delta = 2c \log \left(\frac{r_{\rm g,f}}{r_{\rm g,i}}\right)\,,
\end{equation}
with $r_{\rm g, f}$ and $r_{\rm g,i}$ the final and initial value of the degenerate horizons and the geodesic observer. Therefore, any charged Nariai space-time is at finite distance of its neutral limit. 

%%%%%%%%%%%%%%%%%%%%%%%%%%%%%%%%%%%%%%%%%%%%%%%%%%%%%%%%
\subsubsection{Distance for the RNdS evaporation}
\label{sec:DistRNdSEvap}
%%%%%%%%%%%%%%%%%%%%%%%%%%%%%%%%%%%%%%%%%%%%%%%%%%%%%%%%

\begin{figure}
    \centering
\includegraphics[width=0.8\textwidth]{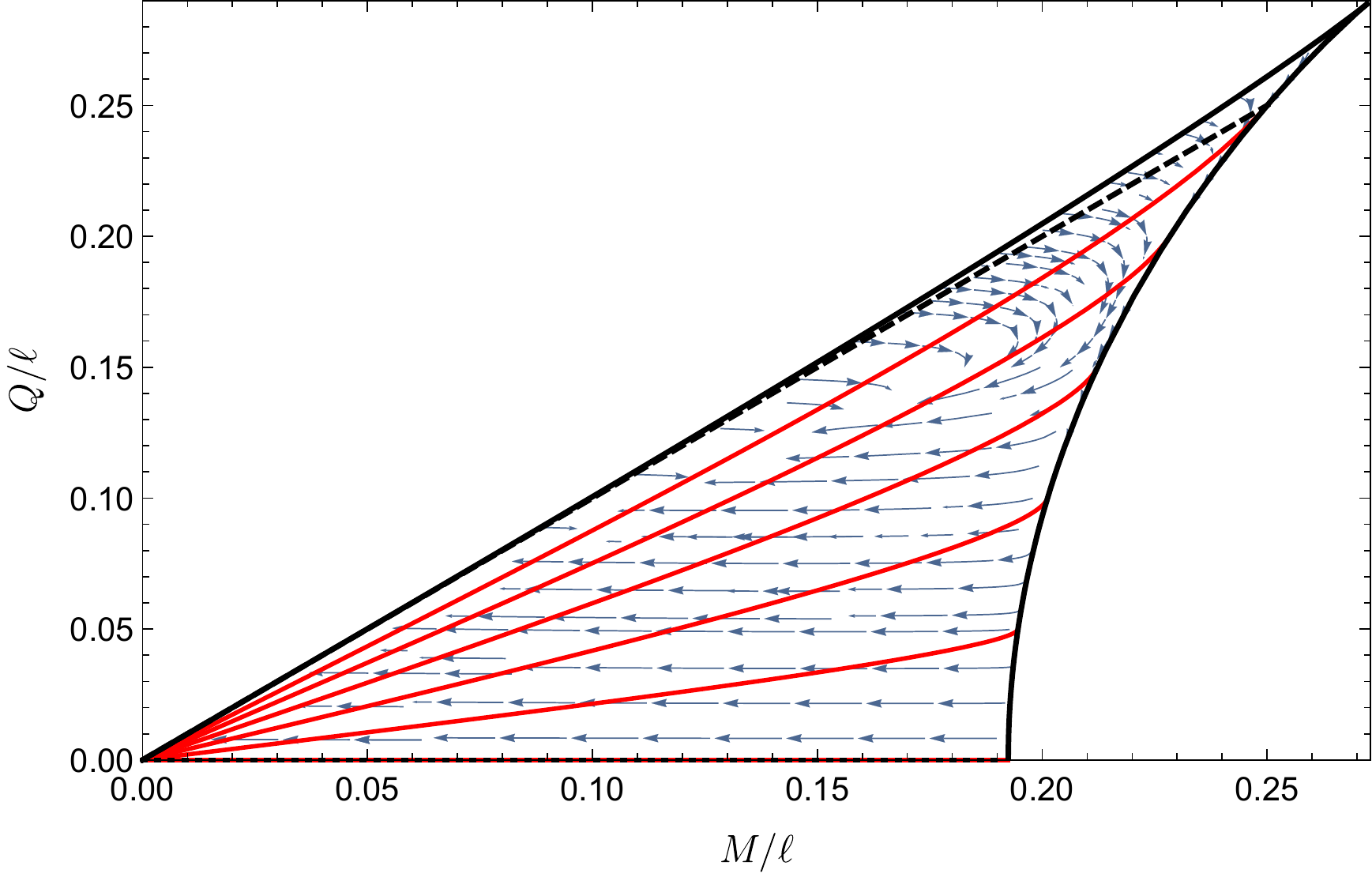}
    \caption{Comparison of the moduli space geodesics (red lines) given by Eqs.~\eqref{eq:xyoflambda} for arbitrary initial conditions and the phase space evaporation in phase space for the family of RNdS space-times (blue arrows). }
    \label{fig:geoVstrajXY}
\end{figure}

Next, we generalize to non-Nariai space-time configurations.
For the evaporation we do not have the analytical expression describing the trajectories in phase space.
However, since the geometric distance~\eqref{eq:distance-final} does only depend on the initial and final points,
we use that empty de Sitter is the final stage of evaporation.
The general solution of the geodesic equation imposes the conditions~\eqref{eq:conditions} for the parametrization $\alpha$ and the mass parameters. Conditions~\eqref{eq:conditions} were already solved in section~\ref{sec:Evolution_Mass_Parameters_Moduli}, leading to Eq.~\eqref{eq:mqloflambda} that determine the dependence of the mass parameters with the moduli space proper time.
In order to fully specify the moduli space geodesic, we still need to choose $\alpha$ such that the observer at $r =\tilde r \alpha$ does not hit a curvature singularity along the evolution. 

The evaporation process can be described in terms of the ratios $x=M/\ell$ and $y=Q/\ell$.
Hence, in order to ease the comparison between the evaporation paths and the projection of the moduli space geodesics, we work in terms of the ratios $x$ and $y$.
These evolve along the flow as 
\begin{align}\label{eq:xyoflambda}
    x(\lambda) = x_{\rm i} (1+c_y \lambda) \sqrt{1+c_\ell \lambda}\,, \quad
    y(\lambda) = y_{\rm i}  \sqrt{1+c_y\lambda} \sqrt{1+c_\ell \lambda}\,.
\end{align}
The endpoint of evaporation is empty de Sitter space, which is arrived at $\lambda = -1/c_y$.
The constant $c_y$ can be absorbed in the definition of the proper time $\lambda$, so the geodesics depend only on one free parameter $c_\ell$.

\begin{figure}
\centering
\includegraphics[width=0.8\textwidth]{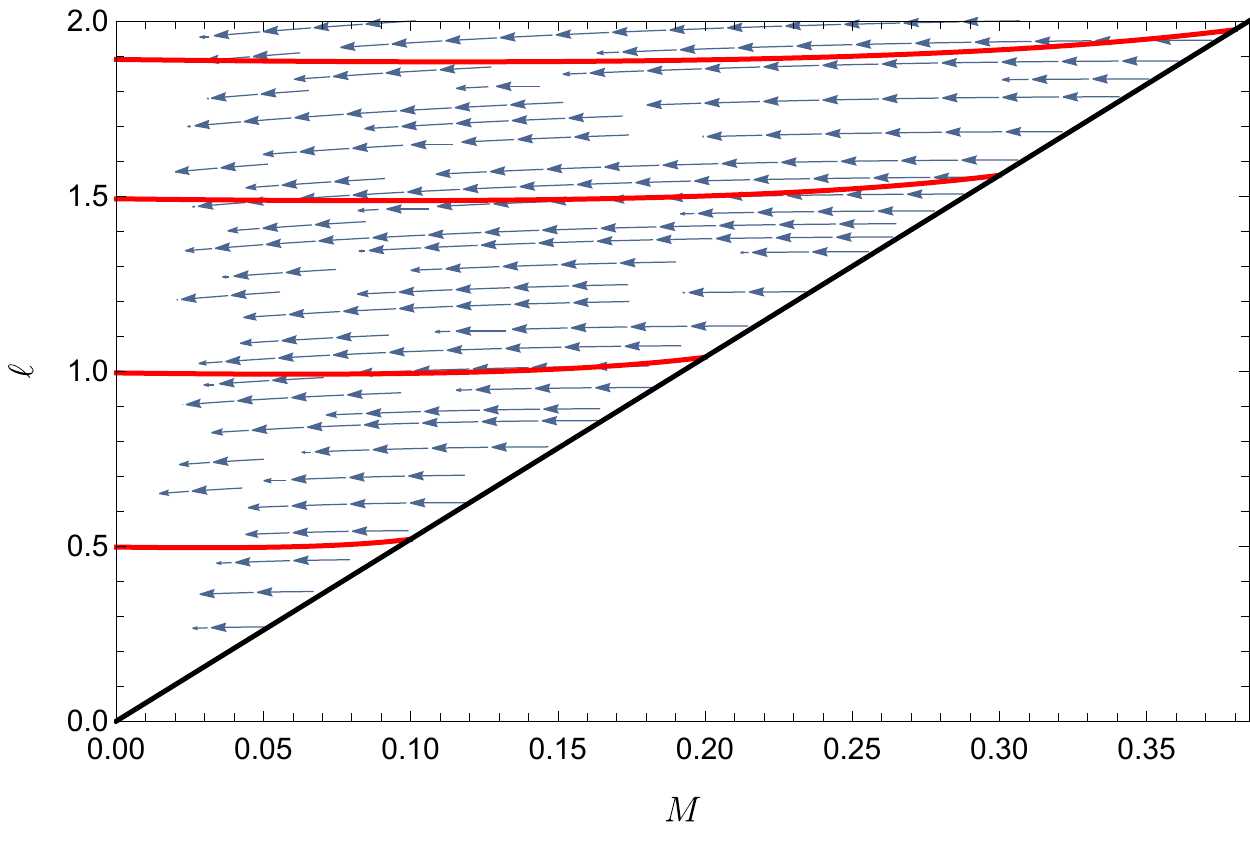}
\caption{Comparison of the moduli space geodesics (red lines) given by Eqs.~\eqref{eq:mqloflambda} for arbitrary initial conditions and the phase space evaporation in phase space for the family of SdS space-times (blue arrows).}
\label{fig:geoVsModSdS}
\end{figure}

The cosmological horizon is defined along the whole evaporation process, so it smoothly interpolates between the initial cosmological horizon $r_{\cc,\rm i} \neq \ell_{\rm i}$ and the de Sitter radius in the pure de Sitter limit $\lim_{\lambda\to 1} r_{\cc} = \ell_{\rm f}$.
Hence, we choose the cosmological horizon as our parametrization $\alpha=r_\cc$\footnote{We fix the free parameter $c_\ell$ through the requirement that $\lim_{\lambda\to 1} r_{\cc} = \ell_{\rm f}$ is satisfied.  }.
The projection of the resulting geodesic paths onto the phase space are represented in Fig.~\ref{fig:geoVstrajXY} by solid red lines for different initial conditions.
As can be seen, these geodesics stay within the physical region $\mathcal{D}$ for the entire interval between the initial and final point of evaporation.
Hence, we can utilize this class of geodesics to infer that the geometric distance between any initial and final point of black hole evaporation is given by
\begin{equation}\label{eq:Dgrc}
\Delta = 4c \log \left(\frac{\ell_{\rm f}}{r_{\cc,\rm i}}\right) = c\,.
\end{equation} Here $r_{\cc,\rm i}$ is the cosmological horizon of the initial space-time and $\ell_{\rm f}$ the cosmological horizon of the final space-time, which is simply given by the de Sitter horizon as the black hole has evaporated completely.
Note that the distance is independent of $c_\ell$.

Different choices of the integration constants of~\eqref{eq:mqloflambda} yield different geodesics parametrized by distinct functions $\alpha$ of the mass parameters. For these other possible families of geodesics we expect that the distance to empty de Sitter space is again given by Eq.~\eqref{eq:Dgrc}.

Next, we discuss the Schwarzschild-de Sitter case. 
Again, the cosmological horizon provides a convenient parametrization, $\alpha = r_\cc$.
It yields smooth, real-valued paths in moduli space along the evaporation trajectories.
In Fig.~\ref{fig:geoVsModSdS}, we show the projection of the moduli space geodesics over the phase space (red lines) together with the evaporation paths (blue arrows) leading to empty de Sitter space.
Notice that in this case the projected geodesics parametrized by $r_\cc$ stay within the physical region.
Similarly to the RNdS case, the evaporation paths do not coincide with the moduli space geodesics.
Nevertheless, they can be used to compute the distance from SdS to dS.
The distance is again finite and explicitly given by Eq.~\eqref{eq:Dgrc}, with $r_{\cc,\rm i}$ the cosmological horizon of the SdS geometry.
The path with parametrization $ \alpha = r_\cc$ is smooth and real-valued for $\lambda\in [0,1]$.

Summarizing, our toy model describes the evaporation of a general RNdS black hole.
The final point of evaporation is empty de Sitter space-time.
For the special case of charged Nariai black holes, our model predicts discharging along the Nariai branch until the neutral Nariai limit is reached. 
In the neutral limit, the model predicts that the black hole losses mass and evaporated towards empty de Sitter space.
By emulating the evaporation trajectories by geodesics in the moduli space, we assigned a geometric distance to the evaporation process.
Our analysis shows that this distance is finite.
For the Nariai case, the geodesics coincide exactly with the evaporation trajectory in phase space.

\section{Summary and conclusions}
\label{sec:Conclusion}
In the first part of the paper, we reviewed to concept of the moduli space of metrics.
Specializing to spherically-symmetric and static space-times, we further refine the prescription of~\cite{Bonnefoy:2019nzv} to compute distances along geodesic paths in the moduli space.
Let $\lambda$ be the proper time of the geodesic path, then we demonstrated that the geometric distance $\Delta$ is always given by
\begin{equation}
    \Delta = c |\lambda_{\rm f} - \lambda_{\rm i}|\,,
\end{equation}
where $c\sim\mathcal O(1)$.
Here, $\lambda_{\rm f,i}$ refer to the proper time at the final and initial point that are connected by a geodesic path, respectively.
This result establishes that space-times are at infinite distance from each other if and only if they are separated by infinite proper time.

Building over this, we moved on to study distances between various space-time configurations that belong to the family of Reissner-Nordstr{\o}m-de Sitter,
which is the most general, spherically symmetric and static space-time.
We find that all space-times with at least one mass parameter are at finite distance from each other.
We further find that the Minkowski limit of infinite mass (and charge) is at infinite distance from every spherically-symmetric and static space-time.
On the other hand, the Minkowski limit, where all mass parameters vanish identically, can be reached at finite distance.
However, due to the freedom to parametrize the geodesics, this same Minkowski limit can also be reached at infinite distance.
In particular, geodesics that represent Weyl rescalings yield infinite distance to both aforementioned Minkowski limits.

As next step, we related the distances to the entropy of the space-time, which we denote by $\mathcal S$.
We first discuss how to define the entropy of multihorizon space-times.
An unambiguous definition is possible only for those space-times that are in thermal equilibrium.
In a thorough case by case study we find that the distance can be related to the entropy as
\begin{equation}
    \Delta \sim 2c \log \mathcal S\,.
\end{equation}
However, for the cases where a multi-horizon space-time flows towards a single-horizon space-time, this relation is valid only asymptotically.
For the cases where a single- or multi-horizon space-time out of thermal equilibrium flows towards the zero-mass Minkowski limit, this relation could not be established.
For all other cases and in particular for Weyl rescalings, the relation holds true.

Finally, we applied our results to black hole evaporation via the combined Hawking and Schwinger effect.
By employing a model based on perturbative and thermodynamic considerations, we describe self-similar evaporation of RNdS black holes.
We find that a general RNdS black hole evaporates towards de Sitter space-time filled with thermal radiation.
The special case of a charged Nariai black hole discharges until it becomes a neutral Nariai black hole.
Combined with our previous results, we find that an evaporating RNdS black hole travels a finite distance to reach its final stage.
We hence expect the entire evaporation process to lie on the landscape.

As a next step, it would be interesting to establish a concept of entropy for the case of multi-horizon space-times that are out of thermal equilibrium.
A precise notion of entropy in these cases will allow to further test the entropy-distance relation.

\acknowledgments
We thank Gia Dvali and Gerben Venken for very useful discussions.
M.L. acknowledges support from a PhD grant from the Max Planck Society.
The work of D.L. is supported by the Origins Excellence Cluster.

\newpage
\appendix
\section{Details on the structure of RNdS space-time}
\label{sec:appA}
In this appendix, we provide some explicit expressions for the RNdS space-time.

The location of the causal horizons are given by the roots of the quartic polynomial $V(r)|_{r=r_h}=0$ with the metric function $V(r)$ given by Eq.~(\ref{eq:V_norm}).
It is straight forward to find that three of the roots are explicitly given by
\begin{align}
\begin{split}\label{eq:rhorizon}
r_- & = \frac{\ell}{2\sqrt{3}} \left(-\sqrt{2+\frac{\rho}{\Theta}+\Theta} + \sqrt{4-\frac{\rho}{\Theta}-\Theta+\frac{12\sqrt{3}M/\ell}{\sqrt{2+\frac{\rho}{\Theta}+\Theta}}}\right)\,,\\ 
r_+ & = \frac{\ell}{2\sqrt{3}} \left(\sqrt{2-\frac{\rho}{\Sigma}-\Sigma} - \sqrt{4+\frac{\rho}{\Sigma}+\Sigma-\frac{12\sqrt{3}M/\ell}{\sqrt{2-\frac{\rho}{\Sigma}-\Sigma}}}\right)\,,\\
r_c & = \frac{\ell}{2\sqrt{3}} \left(\sqrt{2-\frac{\rho}{\Sigma}-\Sigma} + \sqrt{4+\frac{\rho}{\Sigma}+\Sigma-\frac{12\sqrt{3}M/\ell}{\sqrt{2-\frac{\rho}{\Sigma}-\Sigma}}}\right)
\end{split}
\end{align}
where we introduced the short-hand notation
\begin{align*}
\rho &= 1-12\left(\frac{Q}{\ell}\right)^2 \,,\quad
\Sigma = -\left(\lambda + \sqrt{\lambda^2-\rho^3}\right)^{1/3}\,,\\
\Theta &= \left(\lambda-\sqrt{\lambda^2-\rho^3}\right)^{1/3}\,,\quad
\lambda =- 1+54\left(\frac{M}{\ell}\right)^2-36\left(\frac{Q}{\ell}\right)^2\,.
\end{align*}
The fourth root is given by $r_o = -(r_- + r_+ + r_\cc)$.
If the mass parameters are such that the discriminant locus is positive, $D>0$, all four roots are real-valued.
The roots in Eq.~(\ref{eq:rhorizon}) are positive and satisfy $r_\cc > r_+ > r_- > 0$ while the forth root is negative, $r_o<0$, and hence unphysical.
If the discriminant locus is negative, $D<0$, some of the roots are complex-valued, the space-time admits only one causal horizon and exhibits a naked singularity.
Since this case violates the Cosmic Censorship Conjecture, we consider this configuration as unphysical.

Special cases arise when the discriminant locus vanishes, $D=0$.
Two out of the three causal horizons are degenerate.
We can distinguish three cases, which specify further here:
\begin{itemize}
\item
\textit{Extremal space-time:}
In this case, the outer and inner black hole horizon are degenerate, $r_+ = r_ -$.
The degenerate horizons are in thermal equilibrium at zero temperature.
Their mass, charge and de Sitter radius are not independent, but related by the analytic relation
\begin{equation} \label{Eq:extremal}
    \frac{M}{\ell} =\frac{1}{3\sqrt{6}}\sqrt{1+36\left(\frac{Q}{\ell}\right)^2-\rho^{3/2}}\,.
\end{equation}
For fixed values of the cosmological radius $\ell$ the near horizon geometry is $AdS_2\times S^2$.

\item
\textit{Nariai space-time:}
The cosmological and the outer black hole horizon are coincident, $r_c=r_+$.
Both horizons are in thermal equilibrium at zero temperature.
Again, the mass parameters are not independent, but follow the relation
\begin{equation}\label{Eq:Nariai}
    \frac{M}{\ell} =\frac{1}{3\sqrt{6}}\sqrt{1+36\left(\frac{Q}{\ell}\right)^2+\rho^{3/2}}\,.
\end{equation}
The near horizon geometry for a fixed value of the cosmological radius is $dS_2\times S^2$.

\item
\textit{Ultra-cold space-time:}
The intersection of the aforementioned Nariai and extremal space-times is characterized by a triple root of the polynomial equation $D=0$.
The external and Cauchy black hole horizons coincide with the cosmological one, $r_-=r_+=r_\cc$.
All three horizons are in thermal equilibrium at zero temperature.
The ultra-cold space-time is located at
\begin{equation} \label{eq:ultra-cold}
    \frac{M}{\ell} =\frac{1}{3}\sqrt{\frac{2}{3}}\,,\quad \frac{Q}{\ell}=\frac{1}{2\sqrt{3}}\,.
\end{equation}
This can also be seen in Fig.~\ref{fig:sharkfinxy}.
\end{itemize}

\section{Scalar curvature of the moduli space of spherically symmetric and static space-times}
\label{sec:AppB}
The distance conjecture~\cite{Ooguri:2006in} states that the scalar curvature of the moduli space of fields should be strictly negative when the dimension of the moduli space is larger than one.  As detailed in Sec.~\ref{sec:ModuliSpaceAndGeodesics}, we construct the moduli space of spherically symmetric and static space-times, so our moduli space is constructed out of tensor instead of scalar fields. Here we show that the scalar curvature of this moduli space is negative and hence that it is compatible with the statements of the Distance Conjecture~\cite{Ooguri:2006in}. 

From the geodesic equation~\eqref{eq:geodesicEq}, we can extract the Christoffel symbol $\Gamma:\mathcal{M}\times T_g\mathcal{M}\times  T_g\mathcal{M}\to T_g\mathcal{M}$ from the symmetrization of
\begin{equation}
    \ddot{g} = \Gamma(\dot{g},\dot{g})\,,
\end{equation}
thus leading to 
\begin{align*}
    \Gamma_g(h,k)  = &\frac{1}{2}h g^{-1} k  + \frac{1}{2}k g^{-1} h + \frac{1}{4} \text{tr}(g^{-1}hg^{-1}k)g -\frac{1}{4}\text{tr}(g^{-1}h)k -\frac{1}{4}\text{tr}(g^{-1}k)h \\
    & -\frac{1}{4} \langle\text{tr}(g^{-1}hg^{-1}k)\rangle g + \frac{1}{4}\langle\text{tr}(g^{-1}h) \rangle k + \frac{1}{4}\langle\text{tr}(g^{-1}k) \rangle h\,,
\end{align*}where $g\in \mathcal{M}$ and  $h,k\in T^*\mathcal{M}$ in the vector bundle of all symmetric $(0,2)$-tensors on $\mathcal{M}$. 

The curvature of a connection 
\begin{equation}
    R(X,Y)s =( [\nabla_X ,\nabla_Y] -\nabla_{[X,Y]} )s
\end{equation}is called the Riemann curvature tensor for the Levi-Civita connection for the vector fields $X,Y\in \mathfrak{X}(\mathcal{N})$ and a vector field $s:\mathcal{N}\to T\mathcal{M}$  along $f:\mathcal{N}\to\mathcal{M}$. Here $\mathfrak{X}(\mathcal{N})$ denotes the space of smooth vector fields over the smooth manifold $\mathcal{N}$ and $f$ is a smooth mapping.
In local coordinates, the Riemann curvature tensor reads 
\begin{equation}
    R(h,k)l = \dd\Gamma(k)(h,l)-\dd\Gamma(h)(k,l)+\Gamma(h,\Gamma(k,l))-\Gamma(l,\Gamma(h,l))\,.
\end{equation}
The Riemannian curvature for the canonical Riemannian metric on the manifold $\mathcal{M}$ can be computed to be
\begin{align*}
&g^{-1} R_g(h,k)l = - \frac{1}{4} [L,[H,K]] \\
&+ \frac{n}{16} \biggl(H(\text{tr}(KL)-\langle \text{tr}(KL)\rangle-\langle \text{tr}(KL)-\langle \text{tr}(KL)\rangle\rangle )-
 K(\text{tr}(HL)-\langle \text{tr}(HL)\rangle-\langle \text{tr}(HL)-\langle \text{tr}(HL)\rangle\rangle)\biggr)\\
& - \frac{1}{16} \biggl( K(\text{tr}(L)\text{tr}(H)-\langle\text{tr}(L)\rangle\langle\text{tr}(H)\rangle-\langle 2 \text{tr}(L)\text{tr}(H)-\text{tr}(L)\langle\text{tr}(H)\rangle - \text{tr}(H)\langle\text{tr}(L)\rangle ) \\
& \qquad - H(\text{tr}(L)\text{tr}(K)-\langle\text{tr}(L)\rangle\langle\text{tr}(K)\rangle-\langle 2 \text{tr}(L)\text{tr}(K)-\text{tr}(L)\langle\text{tr}(K)\rangle - \text{tr}(K)\langle\text{tr}(L)\rangle ) \biggr) \\
&- \frac{1}{16} \biggl(\left(\text{tr}(HL)-\langle \text{tr}(HL)\rangle\right)\left(\text{tr}(K)-\langle \text{tr}(K)\rangle\right)-\left(\text{tr}(KL)-\langle \text{tr}(KL)\rangle\right)\left(\text{tr}(H)-\langle \text{tr}(H)\rangle\right)\biggr)\text{Id}\,,
\end{align*}
where we have introduced the short-hand notation $H=g^{-1} h$, $K=g^{-1} k$, $L=g^{-1} l$.
Further, $n=\text{dim} M$ is the dimension of $M$.
For isotropic space-times $(M,g)$ the Riemannian curvature tensor simplifies to \begin{equation}\label{eq:RiemCurvIsotropic}
    g^{-1} R_g(h,k)l =  -\frac{1}{4} [L,[H,K]]\,.
\end{equation}

It is not possible to define the Ricci curvature for the manifold $(\mathcal{M},G)$, but a point-wise Ricci-like curvature can be defined~\cite{Gil:1991}.
Taking the point-wise trace of the Riemann curvature tensor~\eqref{eq:RiemCurvIsotropic}, that we denote with the subscript $x$, yields
\begin{equation}
    \text{Ric}_g (h,l)(x) = \text{tr} (k_x \mapsto R_g(h_x,k_x)l_x) = -\frac{1}{4} \left(\text{tr}(H) \text{tr}(L) - n \text{tr}(HL)\right)\,.
\end{equation}
This amounts to compute the point-wise Ricci scalar as
\begin{equation}
    R=-\frac{n}{4} \left(\frac{n(n+1)}{2} -1\right)\,.
\end{equation}
Hence, the manifold $(\mathcal M, G)$ has a negative scalar curvature if the dimension of $M$ is $n\geq 1$.
Note that the scalar curvature is constant and does not depend on moduli space coordinates.

\section{Quasi-static evaporation of RNdS black holes}
\label{sec:appC}
In this appendix, we present technical details of the evaporation model.
Following~\cite{Montero:2019ekk}, we perturb the Einstein equations to linear order, relate the energy-momentum tensor to the Hawking and Schwinger fluxes and extract evolution equations for $M$, $Q$ and $\ell$.
The resulting system of differential equations is under-determined.
By considering the first law of black hole mechanics we supplement the system by an additional differential equation.

%%%%%%%%%%%%%%%%%%%%%%%%%%%%%%%%%%%%%%%%%%%%%%%%%%%%%%%%
%\subsection{Evaporation model for the mass and charge of the RNdS black hole}
\subsection{Perturbative analysis}
\label{sec:EvaporationMass&Charge}
%%%%%%%%%%%%%%%%%%%%%%%%%%%%%%%%%%%%%%%%%%%%%%%%%%%%%%%%

In order to relate the Hawking and Schwinger particle fluxes to the parameters $M$, $Q$ and $\ell$, we follow the analysis of~\cite{Montero:2019ekk} and solve Einstein's equation perturbatively.
We introduce an order $\epsilon$ perturbation in the time and radial components of the RNdS metric (\ref{metricRNdS}) as
\begin{equation}
\label{PerttoMetric}
\delta \dd s^2 = \epsilon (\delta B(\epsilon t,r))\dd t^2 + \delta A(\epsilon t,r))\dd r^2)\,,
\end{equation}
where $\delta A$ and $\delta B$ are arbitrary functions that depend on the radial coordinate and on the ``slow time scale'' $\epsilon t$.
This is the time scale on which the geometry evolves, i.e. $M = M(\epsilon t)$, $Q= Q(\epsilon t)$ and $\ell = \ell(\epsilon t)$.

The change in the geometry is due to the accumulated effect of the Hawking and the Schwinger radiation.
The quantum fluxes source the slowly varying geometry as
\begin{equation}\label{Eq:eom}
(\delta G_{ab}-8\pi G \delta T_{ab}^\text{cl} +\Lambda \delta g_{ab})(\epsilon t) = 8\pi G\delta T_{ab}^\text{q}(t)
\end{equation}
where $\delta T_{ab}^{\text{cl}}(\epsilon t)$ refers to the classical and hence slow variation of the energy momentum tensor and $\delta T_{ab}^{\text{q}} = \langle \delta T_{ab} \rangle $ to quantum perturbations of energy-momentum\footnote{Notice that we have suppressed the term $\epsilon \delta \Lambda(\epsilon t) g_{ab}$ from Eq.~\eqref{Eq:eom}, since we consider the perturbation of this equation to first order i $\epsilon$ and the dependence of the slow time scale of the variation of the cosmological constant gives rise to second order terms in $\epsilon$. }.  
Since both the black hole charge and the background metric vary, the electric field induced by the charged black hole also changes, which we model through an order $\epsilon$ perturbation of the field strength.
\begin{equation} \label{PerttoF}
F\to F+\epsilon\sqrt{\frac{g^2}{4\pi l^2 G}}\delta F(\epsilon t,r) \dd t\wedge \dd r\,.
\end{equation}

Explicitly, the first order perturbation of the metric~\eqref{metricRNdS} results in a first order variation of the Einstein tensor $\delta G_{ab}$, computed by expanding $\delta G_{ab} = \tilde G_{ab} -G_{ab}$ to first order in $\epsilon$, where $\tilde G_{ab}$ is the perturbed Einstein tensor.
The perturbation of the classical energy momentum tensor is computed as $\delta T_{ab}^{\text{cl}} = \tilde T_{ab}^{\text{cl}} -T_{ab}^{\text{cl}}$.
The stress-energy tensor of classical electromagnetism is given by
\begin{equation}\label{T}
 T_{ab}^{\text{cl}} = \frac{1}{4\pi \epsilon_0} \left( F_{a\mu}F^{\mu}_b- \frac{1}{4} g_{ab} F^{\mu\nu}F_{\mu \nu}\right)\,.
\end{equation}
Let us introduce the time-like vector $\xi_{(t)} =\gamma_{(t)} \partial_t$ and space-like vector $v=1/r^2\partial_r$, where $\gamma_{(t)} = V(r_\geo)^{1/2}$ is normalized at the position of the geodesic observer (see Sec.\ref{sec:RNdS}). 
It is convenient to project the linearized Einstein eq.~\eqref{Eq:eom} onto these vectors yielding
\begin{equation}\label{LinEinsteinEq}
\delta_i = 8\pi G\eta_i\,,\quad i=1,\dots 4\,.
\end{equation}
Here, $\delta_i$ correspond to the projection of the classical part as
\begin{equation} \label{PertEinsteinTensor}
\delta G_{ab}-8\pi G \delta T_{ab}^\text{cl} +\Lambda \delta g_{ab} = \delta_1 \xi_a \xi_b + \delta_2 v_a v_b +\delta_3 \xi_{(a}v_{b)} +\delta_4g_{ab}
\end{equation}
to first order in $\epsilon$.
The functions $\delta_i$ are obtained throught the expansion of Eq.~\eqref{Eq:eom} to first order in $\epsilon$, thus yielding
\begin{subequations}
\begin{equation}\begin{split}
\delta_1 = \frac{1}{2 \gamma _t^2}\left(\frac{2 \text{$\delta $A}
   \left(\ell^2 \left(r^2
   \left(3 M^2+M r-r^2\right)-5 M Q^2 r+2
   Q^4\right)+r^4 \left(r (2 r-3 M)+2
   Q^2\right)\right)}{r^8-r^4
   \ell^2 \left(r (r-2
   M)+Q^2\right)} \right.\\
   \left.-\frac{r \text{$\delta
   $B}' \ell^4 \left(r (r-3 M)+2
   Q^2\right)}{\left(r^4-\ell^2
   \left(r (r-2
   M)+Q^2\right)\right){}^2}-\frac{8
   \text{$\delta $F} Q
   \ell^2}{r^4-\ell^2
   \left(r (r-2 M)+Q^2\right)} \right.\\
  \left. -\frac{2
   \text{$\delta $B} \ell^2
   \left(r^4 \ell^2 \left(r (7
   M-2 r)-4 Q^2\right)+\ell^4
   \left(M Q^2 r+M r^2
   (M-r)-Q^4\right)+r^8\right)}{\left(r^4
   -\ell^2 \left(r (r-2
   M)+Q^2\right)\right){}^3}\right.\\
   \left.+\frac{\text{
   $\delta $A}' \left(r (r-3 M)+2
   Q^2\right)}{r^3} +\frac{r^2
   \text{$\delta $B}''
   \ell^2}{r^4-\ell^2
   \left(r (r-2 M)+Q^2\right)}\right)
\end{split}
\end{equation}   
\begin{equation}
\begin{split}
\delta_2 = \frac{1}{2 r^4
   \ell^6}\left(-2 \text{$\delta $A} r
   \left(r^4-\ell^2 \left(r (r-2
   M)+Q^2\right)\right)\times \right.\\
 \left.  \left(-\ell^4 \left(M^2 r-M
   \left(Q^2+3 r^2\right)+2 Q^2
   r+r^3\right)+r^3 \ell^2
   \left(2 Q^2-M r\right)+2 r^7\right) \right.\\
  \left. +8
   \text{$\delta $F} Q r^4
   \ell^4
   \left(r^4-\ell^2 \left(r (r-2
   M)+Q^2\right)\right)+r^6 \text{$\delta
   $B}'' \left(-\ell^4\right)
   \left(r^4-\ell^2 \left(r (r-2
   M)+Q^2\right)\right)\right.\\
  \left. +\frac{2
   \text{$\delta $B} r^4 \ell^4
   \left(r^4 \ell^2 \left(5 M
   r-4 Q^2\right)+\ell^4
   \left(Q^2 r (2 r-5 M)+M r^2 (5 M-3
   r)+Q^4\right)-r^8\right)}{r^4-r_{\text
   {dS}}^2 \left(r (r-2
   M)+Q^2\right)}\right.\\
   \left.+\text{$\delta $B}'
   \left(r^6 \ell^6 (M-r)+2 r^9
   \ell^4\right) -\text{$\delta
   $A}' \left(\ell^2 (M-r)+2
   r^3\right) \left(r^5-r \ell^2
   \left(r (r-2
   M)+Q^2\right)\right){}^2\right)
   \end{split}
\end{equation}  
\begin{equation} 
\delta_3 = \frac{2r}{V(r) \gamma_{(t)}} \dot{V}(r) = -\frac{4}{\gamma_{(t)} V(r)} \left(\dot{M} -\frac{Q}{r} \dot{Q} -\frac{r^3}{\ell^3} \dot{\ell}\right)
\end{equation} \begin{equation}
\begin{split}
\delta_4 =\frac{1}{2} \left(\text{$\delta
   $B}'' +\frac{4
   \text{$\delta $F} Q}{r^2}+\frac{\text{$\delta $B}'
   \ell^2 \left(r (r-3 M)+2
   Q^2\right)}{r^5-r \ell^2
   \left(r (r-2
   M)+Q^2\right)}\right.\\
   \left. -\frac{\text{$\delta
   $A}' \left(\ell^2 (M-r)+2
   r^3\right) \left(r^4-\ell^2
   \left(r (r-2 M)+Q^2\right)\right)}{r^4
   \ell^4}\right.\\
  \left.+\frac{2
   \text{$\delta $B} \left(r^3
   \ell^2 \left(r (7 M-2 r)-5
   Q^2\right)+\ell^4 (r-M)
   \left(Q^2-M r\right)+r^7\right)}{r
   \left(r^4-\ell^2 \left(r (r-2
   M)+Q^2\right)\right){}^2} \right.\\
   \left.   +\frac{2 \text{$\delta
   $A} \left(\frac{r^3 \left(r (4 r-5
   M)+Q^2\right)}{\ell^2}-\frac{
   5 r^7}{\ell^4}+(r-M)
   \left(Q^2-M
   r\right)\right)}{r^5}\right)\,.
   \end{split}
\end{equation} \end{subequations}
The dot denotes the derivative with respect to the slow time scale while the prime stands for the radial derivative of the relevant quantity. Notice that only $\delta_3 \sim\dot{V}(r)$ contains derivative terms with respect to the slow time $\epsilon t$. Consequently, only the liniarized Einstein equation for $\delta_3$ gives rise to a dynamical equation. The functions $\delta_j$ for $j = 1,2,4$  constrain the functions $\delta A, \delta B$ and  $\delta F$.

The terms $\eta_i$ correspond to the projection of the quantum perturbation of the energy-momentum tensor $\delta T_{ab}^{\rm q}$ onto the same vectors,
\begin{equation} \label{PertEnergy}
\delta T_{ab}^{\text{q}}=\eta_1 \xi_a \xi_b + \eta_2 v_a v_b +\eta_3 \xi_{(a}v_{b)} +\delta_4g_{ab}\,.
\end{equation}
It is convenient to define the projection of the perturbed energy momentum tensor on the time-like and space-like vector fields like
\begin{equation}
\mathcal{T} \equiv \delta T_{ab}^{\text{q}} \xi^a v^b, \quad \mathcal{E} \equiv \delta T_{ab}^{\text{q}} \xi^a \xi^b, \quad \mathcal{S} \equiv \delta T_{ab}^{\text{q}} v^av^b, \quad T \equiv g^{ab} \delta
 T_{ab}^{\text{q}}
\end{equation}
Thus, the different $\eta_i$ are the components of the quantum variation of the energy momentum tensor and can be expressed in terms of $\mathcal{T},\mathcal{E},\mathcal{S}$ and $T$ by contracting $\delta T_{ab}^{\text{q}}$ with $\xi^a$ and $v^a$.
Solving the system of linear equations, we obtain
\begin{align}\label{eq:etas}
\begin{split}
\eta_1 &= \frac{r V \gamma_{(t)}^2 \left(T-r^5
   \mathcal{S}\right)+3 \mathcal{E}}{2
   r^2 V \gamma _t^4}\,,\quad
\eta_2 = \frac{1}{2} r^4 \left(3 r^6 \mathcal{S}
   V-r T V-\frac{\mathcal{E}}{\gamma
   _t^2}\right)\,,\\
\eta_3 &=-2\frac{r^4}{\gamma_{(t)}^2}\mathcal{T}\,,\quad
\eta_4 = \frac{1}{2} \left(r^5
   (-\mathcal{S})+\frac{\mathcal{E}}{r V
   \gamma _t^2}+T\right)\,.
\end{split}
\end{align}

We can now extract the first dynamical equation describing the evolution of the mass parameters.
The $i=3$ component of Eq.~\eqref{LinEinsteinEq} reads
\begin{equation}\label{eq:evaporation1:B}
\dot{M} -\frac{Q}{r} \dot{Q} -\frac{r^3}{\ell^3} \dot{\ell}= 4\pi G r^4 \frac{V(r)}{\gamma_{(t)}} \mathcal{T}\,.
\end{equation}
The background parameters $M$, $Q$ and $\ell$ are evolving with the slow time scale $\epsilon t$ so their time derivatives are defined as
\begin{equation*}
\dot{M} = \frac{\dd M}{\dd t}(\epsilon t)|_{t=0}, \quad
\dot{Q} = \frac{\dd Q}{\dd t}(\epsilon t)|_{t=0}, \quad
\dot{\ell} = \frac{\dd \ell}{\dd t}(\epsilon t)|_{t=0}\,.
\end{equation*}

Further, the perturbation to the electromagnetic field strength~\eqref{PerttoF} results in an electric current, because the perturbed electrical current density must satisfy  
\begin{equation} \label{dstarF}
\dd*\tilde F = *\tilde j
\end{equation}
to first order in the perturbation. Expanding Eq.~\eqref{dstarF} to first order in $\epsilon$ yields the evolution of the charge of the black hole sourced by the Schwinger current $\mathcal{J}$
\begin{equation}\label{eq:evaporation2:B}
    \dot{Q} = -4\pi\mathcal{J}\,.
\end{equation} For further details on the derivation of~\eqref{eq:evaporation2:B} we refer to~\cite{Montero:2019ekk}. 

Summarising, the dynamical evolution of the system is dictated by Eqs.~\eqref{Eq:eom} and~\eqref{eq:evaporation2:B} explicitly given by
\begin{equation}\label{eq:evaporation}
    \dot{M} -\frac{Q}{r} \dot{Q} -\frac{r^3}{\ell^3} \dot{\ell} = 4\pi G r^4 \frac{V(r)}{\gamma_{(t)}} \mathcal{T}\,,\quad
    \dot{Q} = -4\pi\mathcal{J}\,.
\end{equation}
Setting $\dot\ell=0$ and $\gamma_{(t)}=1$, we recover the framework of Ref.~\cite{Montero:2019ekk}.

It remains to specify the quantities $\mathcal T$ and $\mathcal J$.
In order to correctly define the quantum fluxes, let us move to a local inertial frame of reference, i.e. to the position of the geodesic observer.
Introducing space-like unit vector $n = \sqrt{V(r_\geo)}\partial_r$, we define
\begin{equation}
    \mathcal{T}_\geo = \delta T_{ab}^\text{q}\xi_{(t)}^a n^b, \quad \mathcal{J}_\geo  = \delta j^a g_{ab} n^b\,.
\end{equation} 
These are related to the previously introduced components as
\begin{equation}\label{eq:fluxesvspert}
  \mathcal{T} = \frac{1}{\sqrt{V(r_\geo)}r^2} \mathcal{T}_\geo, \quad \mathcal{J} = 2\pi \sqrt{G V(r)} r^2 \mathcal{J}_\geo\,.
\end{equation}
Working on the inertial frame defined by the geodesic observer simplifies the computations, while physics remains independent of the coordinate frame.

The mass flux $\mathcal{T}_\geo$ is sourced by the outgoing Hawking radiation from the black hole horizon and the incoming Hawking radiation from the cosmological horizon.
The lack of thermodynamic equilibrium generates a net flux of thermal radiation through the horizon of the black hole that is related to the quantum perturbation of the energy momentum tensor in the direction perpendicular to the generator of the horizon, i.e. the $rt-$component of the quantum metric perturbation.
Also, notice that there is no net flow of charge through the horizon of the black hole, since the same number of outgoing and incoming charge carriers are generated.
Therefore, the term $\mathcal{T}$ only receives a contribution from the Hawking flux.
At the position of the geodesic observer, it is simply given by the Stefan-Boltzmann law
\begin{equation} \label{Eq:Tflux}
  \mathcal{T}_\geo = \sigma (A_\text{c} T_\text{c}^4 - A_+ T_+^4)= \frac{\sigma}{(4\pi)^3}\frac{1}{V(r_\geo)^2}\left(r_\text{c}^2|V'(r_\text{c})|^4-r_+^2|V'(r_+)|^4\right)\,,
\end{equation}
with $\sigma = \frac{\pi^2 k_B^4}{60 c^2 \hbar^3} = \frac{\pi^2}{60}$ Boltzmann's constant in Hubble units.
Here $A_\text{c}$ and $A_+$ as well as $T_\text{c}$ and $T_+$ are the areas and temperatures associated to the cosmological and black hole horizons evaluated at the position of the geodesic observer.
 The red-shift factor $\propto V(r_\geo)$ arises from normalizing the Killing vector field at the position of the geodesic observer Eq.~\eqref{eq:kappaNormalizedrg}.

Finally, we quantify the flux $\mathcal J$, which can be obtained though the integration per unit volume of the Schwinger pair production rate $\Gamma(r)$~\cite{Montero:2019ekk}, thus yielding 
\begin{align}\label{Jflux}
\begin{split}
  \mathcal{J}_\geo = \frac{ G}{ \ell_0^2}\frac{2}{\sqrt{V(r_\geo)} r_\geo^2}\frac{r_\text{c}^2 r_+^2}{r_\text{c}^2+r_+^2} &\int_{S^2} \sin(\theta)\dd\theta  \dd\phi\int_{r_+}^{r_\text{c}} \dd r'r'^2\Gamma(r') \\
  = \frac{ G}{\ell_0^2}\frac{2}{\sqrt{V(r_\geo)} r_\geo^2}\frac{r_\text{c}^2 r_+^2}{r_\text{c}^2+r_+^2} &\frac{q^2Q^2}{\pi^2} \Bigg[\left(\frac{1}{r_+}e^{-\frac{r_+^2}{Q Q_0}}-\frac{1}{r_\text{c}}e^{-\frac{r_\text{c}^2}{Q Q_0}}\right) \\
 &+ \frac{\sqrt{\pi}}{\sqrt{Q Q_0}} \left( \text{Erf}\left(\frac{r_+}{\sqrt{QQ_0}}\right)-  \text{Erf}\left(\frac{r_\text{c}}{\sqrt{Q Q_0}}\right)\right)\Bigg]\,,
\end{split}
\end{align} where $\text{Erf}$ is the error function. Here $\Gamma(r)$ denotes the Schwinger particle production rate in flat space-time
\begin{equation}\label{Eq:PairProductionRate}
   \Gamma(r) = \frac{(qE)^2}{4\pi^3 \hbar^2} \exp\left(-\frac{r^2}{Q Q_0}\right)\,,
\end{equation}
with $E=Q/r^2$ the electric field, $Q_0 = \hbar q / \pi m^2$ and $m$ and $q$ are the mass and charge of the produced pair of particles.
This holds true as long as the charged pair is created in approximately flat space-time, i.e. for $E(r) \gg R(r)$.
This is the case over the whole phase space $\mathcal D$ (due to the assumption of a quasiestatic discharge), except at the neutral line where $E = 0$.
Taking into account the exponential suppression of Eq.~(\ref{Eq:PairProductionRate}), the leading contribution to the current is due to the electron, the lightest charged particle of the Standard Model, so we can approximate $q\approx e$ and $m\approx m_e$ yielding the value of $Q_0$ used in the calculation of the Schwinger flux. 

%%%%%%%%%%%%%%%%%%%%%%%%%%%%%%%%%%%%%%%%%%%%%%%%%%%%%%%%
\subsection{Thermodynamic considerations}
\label{sec:EvapModelsl}
%%%%%%%%%%%%%%%%%%%%%%%%%%%%%%%%%%%%%%%%%%%%%%%%%%%%%%%%

We have found two differential equations for the three parameters $M$, $Q$ and $\ell$.
We thus need one more equation to specify the system.
The first laws for multihorizon space-times as studied in~\cite{Dolan:2013ft} will form the basis for our purpose.

We start with a quick review of~\cite{Dolan:2013ft}.
The cosmological constant contributes a pressure-like term to the first law as
\begin{equation}\label{Eq:P}
P=-\frac{\Lambda}{8\pi} = -\frac{3}{8\pi \ell^2}<0\,.
\end{equation}
The conjugate variable is the thermodynamic volume $V$.
The first law can be written in different versions.
Evaluating the first law at the black hole horizon leads to
\begin{subequations}\label{Eq:FirstLaw}
\begin{equation}\label{Eq:FirstLaw:1}
    \delta M = T_+ \delta S_+ +(\Phi_+-\Phi_\infty) \delta Q+ V_+\delta P \quad\text{ for } r_+<r<\infty\,, 
\end{equation}
while the first law evaluated at the cosmological horizon reads
\begin{equation}\label{Eq:FirstLaw:2}
    \delta M = -T_\cc \delta S_\cc +(\Phi_\cc-\Phi_\infty) \delta Q+ V_\cc\delta P\,, \quad r_\cc<r<\infty\,.
\end{equation}
The first law valid in the bulk between the outer black hole and cosmological horizon then reads
\begin{equation}\label{Eq:FirstLaw:3}
    0 = T_+ \delta S_+ + T_\cc \delta S_\cc +(\Phi_+-\Phi_\cc) \delta Q- V\delta P\,, \quad r_+<r<r_\cc\,.
\end{equation}
\end{subequations}
The subscripts indicate where the thermodynamic potentials are to be evaluated.
Contrary to our discussion in section~\ref{sec:RNdS}, the Killing vector field is normalized to unity at null infinity and hence $\gamma_{(t)}=1$.
Therefore, the thermodynamic potentials appearing in Eq.~\eqref{Eq:FirstLaw} are given by
\begin{align}\label{eq:thermopotentials}
\begin{split}
    \Phi_+ &= \frac{Q}{r_+}\,, \quad  \Phi_\cc = \frac{Q}{r_\text{c}}\,, \quad  \Phi_\infty  = 0\,, \quad
     S_+  = \pi r_+^2\,, \quad S_\cc = \pi r_\cc^2\,, \\
     T_+ &= \frac{1}{4\pi r_+} \left(1-\frac{Q^2}{r_+^2}-3\frac{r_+^2}{\ell^2}\right)\,, \quad T_\text{c} = -\frac{1}{4\pi r_\text{c}} \left(1-\frac{Q^2}{r_\text{c}^2}-3\frac{r_\text{c}^2}{\ell^2}\right) \,,
\end{split}
\end{align}
with $\Phi$ the electric potential, $S=A/4$ the entropy, and $T=|\kappa|/2\pi$ the temperature.
The thermodynamic temperature lacks the red-shift factor with respect to the one appearing in~\eqref{Eq:Tflux}.
Finally, the thermodynamic volumes $V_+$ and $V_\text{c}$ denote the volume comprised between the black hole horizon and infinity and between the cosmological horizon and infinity.
The volume $V$ denotes the volume of the bulk, $V = V_\cc-V_+$.

In the limit of pure de Sitter, Eq. (\ref{Eq:FirstLaw:2}) reduces to
\begin{equation}\label{eq:FirstLawdS}
0 = -T_\text{c} \delta S_\text{c} + V_\text{c} \delta \left(-\frac{\Lambda}{8\pi}\right)\,.
\end{equation}
The cosmological constant has to decrease to respect the second law of black hole mechanics. 
Then, the de Sitter radius evolves according to Stefan-Boltzmann's law\footnote{ Notice that in the pure de Sitter limit $V_\cc = 4\pi \ell^3/3$, so the volume factor $V_\cc$ cancels out with the pressure term when we express it as a function of $\delta \ell$. },
\begin{equation} \label{eq:GibbonsHawkingdS}
    \frac{\dd\ell}{\dd t} = T_\text{c} \frac{\dd S_\text{c}}{\dd t} =\sigma A_\text{c} T_\text{c}^4 = \frac{\sigma}{4\pi^3 \ell^2}\,.
\end{equation}
This implies that a pure de Sitter space-time evaporates to Minkowski.

In the Schwarzschild limit we want to recover the evaporation of the black hole horizon due to the emission of Hawking radiation.
In this limit the first law of thermodynamics~\eqref{Eq:FirstLaw:2}  reads
$\delta M =T_+ \delta S_+$.
The variation of entropy of the black hole horizon is due to the variation of its mass, so we obtain that
\begin{equation}\label{eq:GibbonsHawkingSch}
\frac{\dd M}{\dd t} = T_+\frac{\dd S_+}{\dd t }= -\sigma A_+ T_+^4\,.
\end{equation}

We demand the evolution equation for the de Sitter radius to satisfy the first law~(\ref{Eq:FirstLaw:3}) and to implement the limits discussed above: Gibbons-Hawking radiation of the cosmological horizon in the dS limit and Hawking evaporation of the Schwarzschild black hole.
Hence, for the general RNdS case we plug Eqs.~(\ref{eq:GibbonsHawkingdS}) and~(\ref{eq:GibbonsHawkingSch}) into the first law~\eqref{Eq:FirstLaw:3} valid in the bulk, which yields the following dynamical equation:
\begin{equation}\label{eq:dldt}
    \frac{\dd \ell}{ \dd t} = 4\pi r_\geo^2 \frac{V_\text{dS}}{V_\cc-V_+} \left(V(r_\geo)^2\mathcal{T}_\geo-(\Phi_+-\Phi_\cc)\mathcal{J}_\geo\right)\,,
\end{equation}
where $V_\text{dS} = 4\pi \ell^3/3$ is simply the volume enclosed by the de Sitter radius.
The factor of $4\pi r^2$ accounts for the normalization at the position of a given observer and the term $V(r_\geo)^2$ arises from considering the red-shift factor that appears in Eq.~\eqref{Eq:Tflux} with respect to the temperature definition in Eq.~(\ref{eq:thermopotentials}). Further, we have used that the evolution of the charge of the black hole is exclusively sourced by the Schwinger radiation~\eqref{eq:evaporation2:B}. 

%%%%%%%%%%%%%%%%%%%%%%%%%%%%%%%%%%%%%%%%%%%%%%%%%%%%%%%%
\subsection{Summary}
\label{sec:summary-evaporation}
%%%%%%%%%%%%%%%%%%%%%%%%%%%%%%%%%%%%%%%%%%%%%%%%%%%%%%%%

To summarize,
Eqs.~\eqref{eq:evaporation} and~\eqref{eq:dldt} lead to the following system of differential equations,
\begin{align}
\begin{split}\label{eq:evaporationEqsModel2}
\mathring{M}& = 4\pi r_\geo^2  \Bigg[\left(G \sqrt{V(r_\geo)} + \frac{r_\geo^3 V(r_\geo)^2}{r_\cc^3-r_+^3}\right)\mathcal{T}_\geo -\left(\frac{Q}{r_\geo} + \frac{r_\geo^3}{r_\cc^3-r_+^3} \left(\frac{Q}{r_+} -\frac{Q}{r_\cc}\right) \right) \mathcal{J}_\geo\Bigg]\,,\\
\mathring{Q} &= -4\pi r_\geo^2 \mathcal{J}_\geo\,,\\
\mathring{\ell}  &= 4\pi r_\geo^2\frac{\ell^3}{r_\cc^3-r_+^3} \left(V(r_\geo)^2\mathcal{T}_\geo - \left(\frac{Q}{r_+} -\frac{Q}{r_\cc}\right) \mathcal{J}_\geo\right)\,.
\end{split}
\end{align}
The differential operator $\mathring{ } = \dd/\dd  t_\geo$ denotes the derivative with respect to the proper time of the geodesic observer, which is related to the Schwarzschild time through $\dd t_\geo/\dd t = \sqrt{V(r_\geo)}$.
The expression for the Hawking and Schwinder fluxes are given by Eq.~\eqref{Eq:Tflux} and \eqref{Jflux}, resp.

\bibliographystyle{jhep}
\bibliography{main}

\end{document}